\documentclass[usegraphicx,usenatbib,fleqn]{mn2e}
\usepackage{latexsym}		% to get LASY symbols
\usepackage{graphicx}		% to insert PostScript figures
\usepackage{rotating}           % defines sidways table and figure env.
\usepackage{enumerate}
\usepackage{mathptmx}
\usepackage{amssymb,amsmath}
\usepackage{natbib}
\usepackage{longtable}
\usepackage{pdflscape}
\usepackage{graphics}
\usepackage{setspace}
\usepackage{chngcntr}
\usepackage{hyperref} 
\usepackage{array}
\usepackage{fancyref}
\usepackage[letterpaper,margin=0.8in]{geometry}
\usepackage{subfigure}
\newcommand{\head}[2]{\multicolumn{1}{>{\centering\arraybackslash}p{#1}}{\textbf{#2}}}

\title[Mass Calibration in Stripe 82]{Mass calibration of galaxy clusters at redshift 0.1-1.0 using weak lensing in the Sloan Digital Sky Survey Stripe 82 co-add}
\author[Matthew P. Wiesner, Huan Lin and Marcelle Soares-Santos]{Matthew P. Wiesner $^{1}$\thanks{E-mail:matthewwiesner@aol.com (MPW)}\thanks{Formerly Department of Physics, Northern Illinois University, DeKalb, IL, 60115}, Huan Lin$^{2}$ and Marcelle Soares-Santos$^{2}$  \\
$^{1}$Department of Physics and Astronomy, Purdue University, West Lafayette IN, 47907\\
$^{2}$Fermilab Center for Particle Astrophysics, Fermi National Accelerator Laboratory, P.O. Box 500,  Batavia, IL, 60510}

\begin{document}

\date{Accepted 2015 June 12. Received 2015 May 8; in original form 2015 January 26}

\pagerange{\pageref{firstpage}--\pageref{lastpage}} \pubyear{2014}

\maketitle

\label{firstpage}

\begin{abstract}
We present {galaxy cluster} mass-richness relations found in the Sloan Digital Sky Survey Stripe 82 co-add using clusters found {using a Voronoi tessellation cluster finder}. These relations were found using stacked weak lensing shear observed in a large sample of galaxy clusters. These mass-richness relations are presented for four redshift bins, $0.1 < z \leq 0.4$, $0.4 < z \leq 0.7$, $0.7 < z \leq 1.0$ and $0.1 < z \leq 1.0$. We describe the sample of galaxy clusters and explain how these clusters were found using a Voronoi tessellation cluster finder. We fit an NFW profile to the stacked weak lensing shear signal in redshift and richness bins in order to measure virial mass $(M_{200})$. We describe several effects that can bias weak lensing measurements, including photometric redshift bias, the effect of the central BCG, halo miscentering, photometric redshift uncertainty and foreground galaxy contamination. We present mass-richness relations using richness measure $N_{VT}$ with each of these effects considered separately as well as considered altogether. We also examine redshift evolution of the mass-richness relation. As a result we present measurements of the mass coefficient ($M_{200|20}$) and the power law slope ($\alpha$) for power law fits to the mass and richness values in each of the redshift bins.  We find values of the mass coefficient of {$8.49 \pm 0.526$, $14.1 \pm 1.78$, $30.2 \pm 8.74$ and $9.23 \pm 0.525 \times 10^{13} \; h^{-1} M_{\sun}$} for each of the four redshift bins respectively. We find values of the power law slope of $0.905 \pm 0.0585$, $0.948 \pm 0.100$, $1.33 \pm 0.260$ and $0.883 \pm 0.0500$ respectively. 
\end{abstract}

\begin{keywords}
gravitational lensing: weak, surveys, galaxies: clusters: general.
\end{keywords}

\section{Introduction} \label{zero}
	Many properties of a galaxy cluster can be measured directly, including cluster richness (the number of galaxies in the cluster), the brightness of each of the cluster galaxies and the morphologies of cluster galaxies. However some important quantities cannot be measured directly but must be inferred from measurable properties. Mass is one such quantity. Thus in order to constrain mass we instead measure quantities that can be related to mass; these are called mass proxies. Cluster richness is commonly used as a mass proxy, but in order for it to give meaningful results, the relation between mass and richness must be calibrated. This calibration is referred to as a mass-richness relation. The mass-richness relations presented in this paper are given in four redshift bins, $0.1 < z \leq 0.4$ (low-z), $0.4 < z \leq 0.7$ (mid-z), $0.7 < z \leq 1.0$ (high-z) and $0.1 < z \leq 1.0$ (all-z). The main efforts to develop mass-richness relations previously have focused on clusters at $z < 0.4$.  
	
	Mass-richness relations are important to galaxy cluster cosmology as they make it possible to conduct cluster counts as a function of mass, allowing constraints on cosmological parameters such as $\Omega_M$, $\sigma_8$, $w_0$ and $\Omega_{\Lambda}$ \citep{Vikhlinin14,Rozo09,Zu14}. Higher-redshift (median $z \approx 0.6$) mass-richness relations are especially relevant in the context of current and future large photometric surveys such as the Dark Energy Survey \citep{Diehl14} and the Large Synoptic Survey Telescope \citep{Abate12}. These surveys will assemble large samples of galaxy clusters at redshifts higher than the average $z=0.25$ of previously found mass-richness relations. Beyond cosmology, measurements of cluster mass at higher redshift will provide knowledge about the evolution of galaxy clusters with redshift and about the distribution and mass of dark matter halos \citep{Kauffmann95, Andreon14}.  
	
	Methods of determining cluster mass include X-ray temperature \citep{Ettori02}, cluster velocity dispersion \citep{Allen11}, frequency bias caused by the Sunyaev-Zeldovich effect \citep{Ade13} and tangential shear caused by weak lensing \citep{Schneider}. (For more background on lensing theory, see especially \citet{Schneider}, \citet{Narayan} or \citet{Mollerach}.)  Although weak lensing signal can be minimal in single clusters, it can be maximized by using stacked shear measurements, combining shear signal in bins of similar richness and redshift. 
	
	We observed stacked weak lensing shear as well as cluster richness in the Sloan Digital Sky Survey Stripe 82 co-add using a sample of galaxy clusters found using the Voronoi Tesselation method \citep{Soares10}. The Stripe 82 co-add reaches to higher redshift (median $\approx 0.6$) than previous surveys such as the Sloan Digital Sky Survey main sample (median $z=0.25$). The mass-richness relation with the best statistics to date \citep{Johnston07} was based on the maxBCG cluster catalog \citep{Koester07b} of clusters found in the SDSS main sample.
	
	In $\S$ \ref{one} we summarize previous measurements of mass-richness relations. In $\S$ \ref{two} we describe the properties of our data set. In $\S$ \ref{four} we describe how stacked weak lensing shear measurements were done and how mass was found from shear results. In $\S$ \ref{five} we describe systematics that introduced uncertainties into our mass measurements and how we compensated for them. In this section we also describe tests conducted to verify the strength of the tangential shear signal. In $\S$ \ref{six} we present our results for mass-richness relations for the Stripe 82 co-add including each of the systematics individually and then altogether. Finally in $\S$ \ref{seven} we present an initial analysis of the redshift evolution of the mass-richness relation. In this paper we take $\Omega_M=0.3$, $\Omega_{\Lambda}=0.7$, $H_0=100h$ and $h=1$.  
	\section{Previous Measurements} \label{one}
	\citet{Johnston07} developed a mass-richness relation by using average shear profiles from stacked weak lensing measurements around 130,000 galaxy clusters at median redshift of $0.25$ found in the SDSS \citep{Johnston07,Sheldon07}. These clusters were taken from the maxBCG cluster catalog \citep{Koester07b}. For richness, they measured $N_{200}$, the number of cluster galaxies inside radius $r_{200}$ \citep{Hansen05}. To find cluster mass they measured stacked weak lensing shear in a set of richness bins. They then fit a model to these shear profiles. There are five terms in their ultimate model of the shear data, including (1) the BCG point mass; (2) the NFW profile; (3) the mean surface mass profile for miscentered clusters (i.e., those not centered on the BCG); (4) the mass of neighboring halos (the two halo term); and (5) the non-linear shear effect. After finding cluster masses ($M_{200}$) as a function of richness, Johnston et al. fit a power relation to their results, of the form
\begin{equation}
M_{200}=M_{200|20}\left( \frac{N_{200}}{20}  \right)^{\alpha}
\end{equation}  
with a mass coefficient $M_{200|20}$ describing the cluster mass at $N_{200}=20$ and a power law slope of $\alpha$. Their final mass-richness relation is
\begin{equation}
\begin{small}
M_{200}=(8.8 \pm 1.2 \times 10^{13} \; h^{-1} M_{\sun})\left(\frac{N_{200}}{20}\right)^{(1.28 \pm 0.04)} 
\end{small}
\end{equation}

	More recently a mass-richness relation was measured in the SDSS Stripe 82 co-add \citep{Simet11}. The clusters, being a subset of the clusters used in Johnston et al., have the same median redshift of $z \approx 0.25$, but many more source galaxies can be observed as the Stripe 82 co-add can reach magnitude 23 in i-band (with 50\% completeness). Simet et al. consider several systematics, including error introduced by treating the brightest cluster galaxy (BCG) as being in the center of the halo when it is not. This is called halo miscentering. They consider miscentering of the BCG by creating a set of mock catalogs from their data. These catalogs preserve galaxy positions, shape errors, photometric redshifts and more but replace actual source galaxy shears with expected shears from a shear model. Since the shear model is produced by a halo model, they control how many of the halos have a miscentered BCG and can obtain information on the actual halo masses. They then fit an NFW profile to the mock data and find that the measured mass is underestimated due to miscentering of the BCG. They then find a relationship between the richness and the scale of miscentering:
\begin{equation}  \label{simet_miscenter}
\frac{M_{200,true}}{M_{200,mis}}=1.44\pm0.17\left(\frac{N_{200}}{20}\right)^{-0.21\pm 0.18}
\end{equation}
They also find a mass-richness relation of
\begin{equation}
M_{200}=(9.56 \pm 0.75 \times 10^{13} M_{\sun})\left(\frac{N_{200}}{20}\right)^{(1.10 \pm 0.12)} 
\end{equation}

	\citet{Ford14a} describe measurements of lensing magnification in the Canada-France-Hawaii Telescope Lensing Survey (CFHTLenS) using cluster candidates found using the 3D-Matched-Filter cluster-finder of \citet{Milk}. In another analysis of the same data \citet{Ford14b} describe measurements of weak lensing shear in the CFHTLenS. Measurements are made for more than 18,000 cluster candidates at $0.2 \leq z \leq 0.9$. In the first paper the authors fit NFW profiles to the lensing magnification measurements and use these to find best-fitting values for $M_{200}$. These values are then used to produce a mass-richness relation. Systematics considered include cluster miscentering, photometric redshift errors and the two-halo term (that is dark matter structures nearby the cluster). The final mass-richness relation obtained is
\begin{equation}
 M_{200}=(2.2 \pm 0.2 \times 10^{13} M_{\sun})\left(\frac{N_{200}}{20}\right)^{(1.5 \pm 0.1)} 
\end{equation}
In the second paper, the authors stack weak lensing shear signal, fit the result to an NFW profile and correct for miscentering. They find a final mass-richness relation of     
\begin{equation}
 M_{200}=(2.7_{-0.4}^{+0.5} \times 10^{13} M_{\sun})\left(\frac{N_{200}}{20}\right)^{(1.4 \pm 0.1)} 
\end{equation}
	
	\citet{Reyes09} describe a study of $\approx 13,000$ clusters from the maxBCG catalog at $0.1 \leq z \leq 0.3$. While this is the same sample as that presented by \citet{Johnston07}, their approach is different in that they try different combinations of mass tracers to identify the optimal mass tracer for galaxy clusters. They find cluster masses using stacked weak lensing shear measurements. They consider systematics including photometric redshift errors, cluster miscentering, contamination of shear signal and intrinsic alignment of source galaxies. They consider the reliability of several mass proxies, including richness ($N_{200}$), cluster luminosity ($L_{200}$) and BCG luminosity ($L_{BCG})$, ultimately combining $N_{200}$ and $L_{BCG}$ into a mass-proxy relation. Along the way they provide a mass-richness relation of
\begin{equation}
 M_{200}=(1.42 \pm 0.08 \times 10^{14} M_{\sun})\left(\frac{N_{200}}{20}\right)^{(1.16 \pm 0.09)} 
\end{equation} 
In Section \ref{others} we compare our results to the previous results discussed in this section. 

\section{Data}  \label{two}
	\subsection{Stripe 82}
	The Sloan Digital Sky Survey (SDSS) began in 1998 and sought to image 10,000 $deg^2$ in the North Galactic Cap and to take spectroscopy of one million galaxies and one hundred thousand quasars in this same region \citep{Annis}. The SDSS uses a 2.5-m telescope located at Apache Point Observatory. The SDSS camera has 24 2048x2048 CCDs of pixel scale $0.396 $ arcsec. Stripe 82 is a section of the footprint of the Sloan Digital Sky Survey (SDSS); Stripe 82 falls along the celestial equator between declinations of $-1.25^{\circ}$ and $+1.25^{\circ}$. Stripe 82 was imaged repeatedly during the fall when the North Galactic Cap was not observable. The main SDSS area reaches $r\approx22.4$ and median seeing of $1.4 $ arcsec, while Stripe 82 reaches about 2 magnitudes deeper and achieved median seeing of $1.1 $ arcsec. This is possible as about 20 runs were taken of each field in the Stripe 82 co-add and these data were then co-added.
	
\subsection{Cluster Samples}
	The maxBCG cluster catalog \citep{Koester07a} is a catalog of 13 823 clusters found in the Sloan Digital Sky Survey main survey using the maxBCG method \citep{Koester07b}. The maxBCG method involves searching for clusters based on the presence of a brightest cluster galaxy (BCG) and a number of E/S0 ridgeline galaxies nearby of similar color. The maxBCG clusters were found in a region of area $7500 \; deg^2$ and have $0.1 \leq z \leq 0.3$. Values of $N_{200}$ range from 10-190. We used the maxBCG catalog as a control sample at low z, verifying that our measurements of weak lensing shear reproduced the mass-richness relation previously found from the maxBCG full sample \citep{Johnston07}. We also used the maxBCG cluster sample to find a relation between $N_{200}$ and our richness measure $N_{VT}$ and we utilized a correction for halo miscentering based on maxBCG cluster data. 

\begin{figure} 
\begin{center}
\includegraphics[scale=0.3, angle=90]{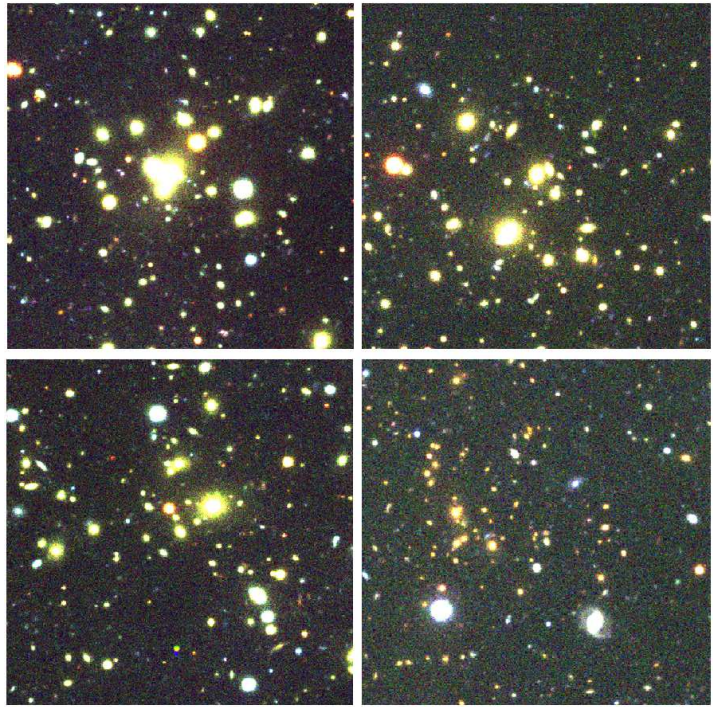}
\caption{Images of the central regions of four of the higher-richness clusters from the new Voronoi tessellation cluster sample in the Stripe 82 co-add. These clusters (proceeding clockwise from the top left) are at (319.70833, 0.55727617), (328.61234, 0.64274948), (16.227354, 0.066838292), and (345.63669, 0.040305327). They have $N_{VT}$ of 99, 78, 127 and 76 and $z$ of 0.28, 0.33, 0.30 and 0.69 respectively. Images are taken from the SDSS DR7 Stripe 82 Data Archive Server at \url{http://das.sdss.org/www/html/imaging/dr-75.html}. Each image has size of approximately 2.5 arcmin square.}  \label{fourclust}
\end{center}
\end{figure}

	We produced a {catalog of 19 316 clusters} from the Stripe 82 co-add using a Voronoi tessellation (VT) cluster finder \citep{Soares10}. A sample of four higher-richness clusters from the sample are shown in Figure \ref{fourclust}. 
A description of the development and the properties of this cluster catalog is in preparation \citep{Soares14}. Voronoi tessellation is a method of relating distances between different locations in a large array of points (called \textit{seeds}). In a Voronoi tessellation, regions called Voronoi cells are defined in which all points within that cell are closer to that seed than to any other seed. In this method, the seeds would be galaxies. If the number and size of Voronoi cells exceed certain criteria, then this region is declared to be a cluster. This was found to be a way of assembling a cluster sample with high purity and completeness while not depending on galaxy magnitude or color \citep{Soares10}. The total number of cluster members is defined by the number of objects within the overdense region in the VT method. Thus the richness measure is not $N_{200}$ but is instead a new richness measure called $N_{VT}$. 

\begin{figure}  
%\begin{center}
\includegraphics[scale=0.35, angle=90]{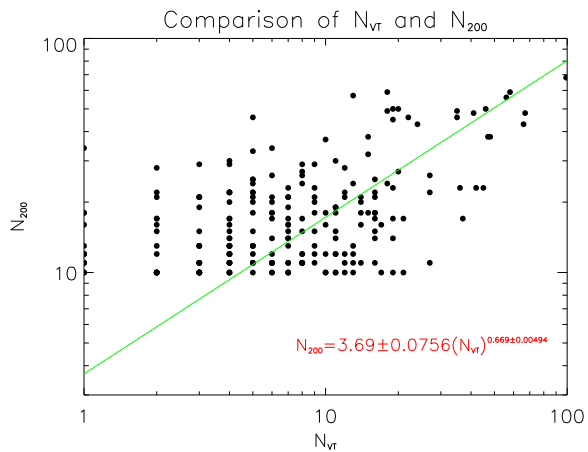}
\caption[$N_{VT}$ vs $N_{200}$.]{{A plot of $N_{VT}$ vs. $N_{200}$} for 230 galaxy clusters matched between the maxBCG cluster sample and our VT cluster sample.} \label{VTvs200}
%\end{center}
\end{figure}

	We wanted to find an approximate relation between $N_{200}$ and $N_{VT}$ both to compare results for each and in order to use Equation \ref{simet_miscenter} for miscentering corrections (this equation depends on $N_{200}$). In order to do this, we found galaxy clusters listed both in the maxBCG catalog and in the Stripe 82 VT catalog. We did this by searching the maxBCG catalog for clusters {within $1.0 $ arcmin} of the cluster center in the VT catalog. We found {230 galaxy clusters} in both catalogs using this method. We plotted the $N_{200}$ from the maxBCG catalog against the $N_{VT}$ from the VT catalog; the {result is shown in Figure \ref{VTvs200}.}
There is not a clear relation between $N_{VT}$ and $N_{200}$. Nevertheless we tried to find an approximate relation by fitting a power law to the data (the power relation
was a better fit than a simple linear fit). The equation we obtained relating the two richness measures was:  
%I used rich_compare.pro to make this and the plot.
\begin{equation}  \label{convertNVT}
N_{200}=3.69 \pm 0.0756(N_{VT})^{0.669 \pm 0.00494}
\end{equation}  

\begin{figure} 
\begin{center}
\includegraphics[scale=0.35, angle=90]{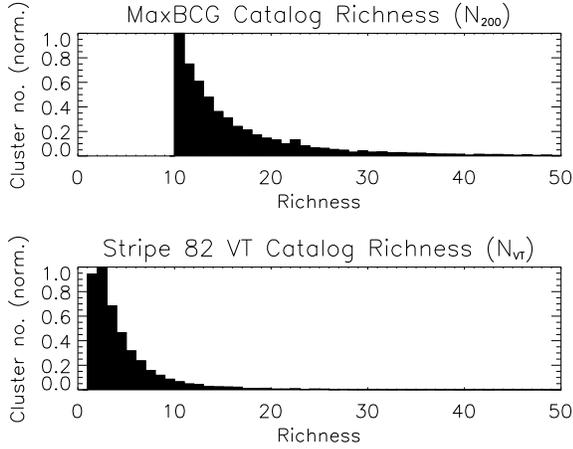}
\caption{Histograms of richness for the maxBCG and the Stripe 82 VT cluster catalogs. Histograms are normalized by their respective maximum values.}  \label{rich_hist}
\end{center}
\end{figure}

\begin{figure} 
\begin{center}
\includegraphics[scale=0.35, angle=90]{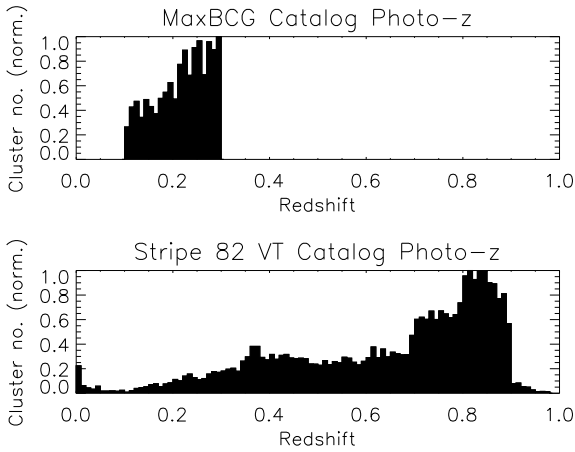}
\caption{Histograms of redshift for the maxBCG and the Stripe 82 VT cluster catalogs. Histograms are normalized by their respective maximum values. The peak at high redshift is due to limitations in the photometric redshifts of the input galaxy catalog.}  \label{red_hist}
\end{center}
\end{figure}

	Finally we consider the richness and redshift ranges of the galaxy clusters, where the redshift used is the photometric redshift found using a neural network algorithm. The photo-z used is that of the central galaxy in the cluster, the galaxy of highest local density as determined by the VT cluster finder. For the 19 316 Stripe 82 co-add clusters the redshift range is $0 \leq z \leq 0.98$ and the richness range {is $1 \leq N_{VT} \leq 99$}. For the maxBCG catalog the redshift range is $0.1 \leq z \leq 0.3$ and the richness range is $10 \leq N_{200} \leq 188$. In our analysis of the VT clusters, we imposed a criterion that clusters needed to be at $z=0.1$ or higher (before any photo-z corrections) because as redshift approaches 0, the significance of the $N_{VT}$ detection decreases significantly. In Figures \ref{rich_hist} and \ref{red_hist} we present histograms of richness and redshift for the maxBCG and the Stripe 82 co-add cluster samples.

Source galaxies were selected from a catalog of galaxies found in the Stripe 82 co-add. In order to produce this catalog, cuts were made on magnitude, size and ellipticity components \citep{Lin12}. There are 5875 133 objects in this catalog, ranging from $0.3 \leq z_{phot} < 1.60$ and i-band magnitude $18 < i < 24$. All galaxies in this catalog include shape information measured by the SDSS pipeline. This catalog also contains photometric redshifts measured for each galaxy using the artificial neural network technique \citep{Reis12}. These photo-zs were trained using a sample of about 83 000 galaxies that have spectroscopic redshifts measured for them.  
\label{lastpage}

\section{Shear Measurements}  \label{four}
	Weak lensing can cause a systematic change in the ellipticities of many galaxies behind the galaxy clusters; this observed change is called \textit{shear}, represented by $\gamma$. Shear has two components, tangential ($\gamma_t$) and orthotangential or cross shear ($\gamma_x$), given by \citep{Schneider}
\begin{equation}
\begin{array}{l}
\gamma_t=-{\Re} \left[\gamma e^{-2i\phi}   \right]
\\
\gamma_x=-{\Im} \left[\gamma e^{-2i\phi}   \right]
\end{array}
\end{equation}
where $\phi$ is the position angle measured between the center of the cluster and the lensed galaxy. Only tangential shear is produced by weak lensing, so we expect no signal in orthotangential shear. Since the observed tangential shear is related to cluster mass, weak lensing can provide a measure of cluster mass. Tangential shear is found for a combination of galaxy clusters and source galaxies, where source galaxies are galaxies nearby clusters in projection, but at a larger redshift than the clusters. The source galaxies are the galaxies whose shape is measured, and the mass associated with the clusters is the cause of the tangential shear that describes the shape variation. 

	The quantity that can be directly observed in images is galaxy ellipticity, which we found by measuring second order moments of galaxies. In order to convert this to shear, we divided by responsivity ($R$). For the Stripe 82 co-add, we used 
\begin{equation}
R=2(1-\sigma^2_{SN})=1.73
\end{equation}
where $\sigma_{SN}$ is the intrinsic galaxy shape noise in the SDSS, taken to be $\sigma_{SN}=0.37$ \citep{Hirata04}. Shape noise describes the intrinsic variation in galaxy shapes, independent of any measurement uncertainty. For Stripe 82 data, PSF deconvolution was done by R. Reis using the Hirata-Seljak method \citep{Hirata03}. 

	Even in the Stripe 82 co-add data, {which is much deeper than the SDSS main sample}, {the} weak lensing signal was likely to be noisy for an individual cluster. Thus in order to maximize shear signal to noise we measured stacked weak lensing shear (following \citet{Johnston07}). To stack weak lensing shear we measured shear for all source galaxies found in a particular richness and redshift bin. Thus we took the average of the source galaxy shears in that richness/redshift bin at each radius bin. For each bin we had a set of 30 measurements of average shear for 30 increments of distance from the central BCGs (from $0-3.0 \; h^{-1}$ Mpc in steps of 0.1).

	Tangential shear is related to the mass and concentration of the lensing mass distribution. In order to find these quantities, we apply an NFW profile and find the best-fitting mass and concentration that produce the observed shear \citep{Wright00}. Weak lensing shear as a function of radius predicted by the NFW model is given as 
\begin{equation} \label{gammagamma}
\gamma_{NFW}(x)=\frac{\overline{\Sigma}_{NFW}(x)-\Sigma_{NFW}(x)}{\Sigma_{crit}}
\end{equation}
Here $x$ is a dimensionless radius equal to $r/r_s$, where $r$ is the distance from the cluster center in the lens plane and $r_s$ is the scale radius in the NFW model, the radius at which the density changes from a $1/r$ relation to a $1/r^3$ relation. $\Sigma_{NFW}(x)$ is the surface mass density of the galaxy cluster's dark matter halo in the NFW model, $\overline{\Sigma}_{NFW}(x)$ is the mean surface mass density of the halo and $\Sigma_{crit}$ is
\begin{equation}
\Sigma_{crit}=\frac{c^2 d_{S}}{4 \pi G d_{L} d_{LS}}
\end{equation}
Here $d_{S}$, $d_{L}$ and $d_{LS}$ are angular diameter distances to the source galaxy, to the lensing cluster and from lensing cluster to source galaxy, respectively. We implemented the series of relations in \citet{Wright00} predicting shear as a function of radius in the NFW model. We then fit observed shear to this model in order to find the {mass ($M_{200}$) that would lead to} the measured values of tangential shear in our richness and redshift bins. {We held $c_{200}$ constant while} doing these fits, as we found that if we allowed both $M_{200}$ and $c_{200}$ to float simultaneously, values for $c_{200}$ were lower than would be expected and were not well constrained. We chose values of $c_{200}$ by using the concentration-mass relation of \citet{Duffy08}, given as:
\begin{equation}
c_{200}(M,z)=A(M/M_{pivot})^B(1+z)^C
\end{equation}
We use the values given for the full (not only relaxed) cluster sample and redshift range $0-2$. For this sample, $A=5.74 \pm 0.12$, $B=-0.084 \pm 0.006$, $C=-0.47 \pm 0.04$ and $M_{pivot}=2\times10^{12} \; h^{-1} M_{\sun}$. We found $c_{200}$ in the range of $10^{12}-10^{15} \; h^{-1} M_{\sun}$ and then took the average value of $c_{200}$ across this range of values for $M_{200}$ and for the median redshift value in each redshift bin (0.25, 0.55, 0.85 and 0.55 for the low-z, mid-z, high-z and all-z bins, respectively). 

\begin{figure}
\centering
\includegraphics[scale=0.41, angle=90]{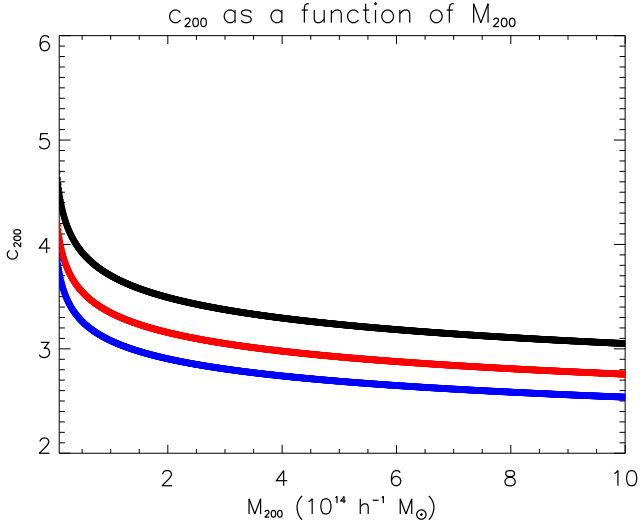}
\caption[A plot of $c_{200}$ as a function of $M_{200}$.]{{A plot of $c_{200}$} as a function of $M_{200}$ using the relationship described in \citet{Duffy08}. The uppermost (black) curve is for median z of 0.25, the middle (red) curve is for median z of 0.55 and the lowest (blue) curve is for median z of 0.85.}  \label{cfuncm}
\end{figure}
We found $c_{200}=3.33$ for the low-z bin, $c_{200}=3.01$ for the mid-z and all-z bins and $c_{200}=2.77$ for the high-z bin. In Figure \ref{cfuncm} we plot $c_{200}$ as a function of $M_{200}$ for the three median redshifts.

	We then calculated a scale factor to scale all cluster shears to the same redshift. This was done because weak lensing shear is a function of redshift, and while all clusters in a redshift/richness bin are at similar redshift, they are not at identical redshift. The scale factor $S$ is
\begin{equation} \label{scaleslifted}
S=\frac{d_{S} d_{L\_fid} d_{LS\_fid}}{d_{S\_fid} d_{L} d_{LS}}
\end{equation}
The ones marked "fid" mean at the fiducial redshift, which is $z_{fiducial\_lens}=0.55 $ for the lens and $z_{fiducial\_source}=0.75$ for the source. Thus all objects are scaled to these two redshifts, which were chosen as approximately the mean redshifts for the cluster and the source galaxies respectively. Equation \ref{scaleslifted} was obtained by taking the ratio of $\Sigma_{crit}$ for the cluster redshift to that the fiducial redshift. We calculate average tangential shear by using inverse variance weighting in order to overcome the effects of large values of $S$ that arise when $d_{L}$ and $d_{S}$ are close in value. 

\begin{figure}
\centering
\includegraphics[scale=0.35, angle=90]{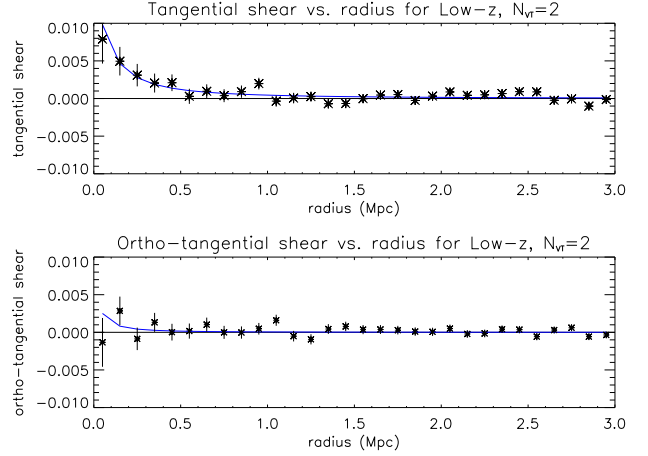}
\caption[Sample shear profiles for Stripe 82 co-add.]{A sample shear profile. This profile is for Stripe 82 co-add, low-z bin ($0.1 \leq z \leq 0.4$), $N_{VT}=2$ with no corrections for systematics applied. Note that the upper tangential shear profile shows a shear signal while the lower orthotangential shear profile does not. There were 614 clusters in this bin and 2 699 077 source galaxies.}  \label{sampleplot}
\end{figure}

	Finally we take the measured average shear values in each bin in the range of $0-3.0 \; h^{-1}$ Mpc from the central BCGs and the predicted shear values found from Equation \ref{gammagamma} and minimize the $\chi^2$ function relating them. {The value of distance from the cluster center} in each bin was the midpoint of the distance bin; for example, in the distance range of $0.1-0.2 \; h^{-1}$ Mpc, the distance used for fitting was $0.15 \; h^{-1}$ Mpc. The values of $M_{200}$ that minimize the $\chi^2$ are taken as the NFW model fits for cluster mass. A sample plot of shear profiles with best-fitting curves is shown in Figure \ref{sampleplot} for the low redshift ($0.1-0.4$) bin with $N_{VT}=2$.  
Note that there is good evidence for a tangential shear signal, but orthotangential shear is consistent with zero. 
	
	In Tables \ref{allresults1} and \ref{allresults2} we present measurements of richness and stacked cluster mass as well as reduced $\chi^2$ for a fit to 0 for each richness and redshift bin. 
\begin{table}
\caption[Results of weak lensing shear fits.]{Weak lensing shear fit results for low-z ($0.1-0.4$) and mid-z ($0.4-0.7$) clusters. This table includes values for richness ($N_{VT}$), mass ($M_{200}$, in units of $10^{14} \; h^{-1} \; M_{\sun}$) and null test (fit to 0) reduced $\chi^2$. Values for concentration ($c_{200}$) are $3.33$ for low-z and $3.01$ for mid-z. No corrections for systematics are applied to obtain these values.} \label{allresults1}
\begin{center}
\begin{tabular}{c c c}
{$N_{VT}$}&{$M_{200}$}&$\chi_{red}^2$ \\    
LOW-Z&&  \\
1  $\pm$ 0 &  0.0169 $\pm$  0.0100 &  1.57  \\
2  $\pm$ 0 & 0.0471 $\pm$ 0.0135 &  1.84   \\
3  $\pm$ 0 & 0.0797 $\pm$ 0.0200 &  2.37   \\
4  $\pm$ 0 & 0.0695 $\pm$ 0.0238 &  1.66  \\
5  $\pm$ 0 & 0.131 $\pm$ 0.0321 &  2.32   \\
6 $\pm$  0 & 0.117 $\pm$ 0.0361 &  1.67   \\
7 $\pm$ 0 & 0.181 $\pm$ 0.0510 &  1.98   \\
8 $\pm$ 0 & 0.130 $\pm$ 0.0496 &  1.97   \\
9 $\pm$ 0.500 & 0.207 $\pm$ 0.0478 &  2.32   \\
11 $\pm$ 0.500 & 0.464 $\pm$  0.0773 &  4.26   \\
14 $\pm$ 1.00 & 0.377 $\pm$ 0.0749 &  3.91  \\
18 $\pm$ 2.00 & 0.288 $\pm$ 0.0706 &  2.42   \\
23 $\pm$ 4.50 & 0.421 $\pm$ 0.0973 &  2.68   \\
36 $\pm$ 4.50 & 0.904 $\pm$ 0.204 &  2.62  \\
46 $\pm$ 4.50 & 1.21 $\pm$ 0.258 &  2.49   \\
54 $\pm$ 4.50 & 1.92 $\pm$ 0.532 &  2.10  \\
66 $\pm$ 3.00 & 1.69 $\pm$ 0.536 &  2.15   \\
78 $\pm$ 4.00 & 2.081 $\pm$ 0.870 &  1.81  \\
99 $\pm$ 0 & 0.560 $\pm$ 0.842 &  1.33  \\ \\
MID-Z&&  \\
1 $\pm$ 0 & 0.0614 $\pm$ 0.0229 & 1.49 \\
2 $\pm$ 0 & 0.00693 $\pm$ 0.00991 & 1.55 \\
3 $\pm$ 0 & 0.0486 $\pm$ 0.0240 & 1.23 \\
4 $\pm$ 0 & 0.173 $\pm$ 0.0459 & 1.91 \\
5 $\pm$ 0 & 0.130 $\pm$ 0.0526 & 2.084 \\
6 $\pm$ 0 & 0.217 $\pm$ 0.0708 & 2.016 \\
7 $\pm$ 0 & 0.204 $\pm$ 0.0802 & 1.44 \\
8 $\pm$ 0 & 0.506 $\pm$ 0.132 & 1.91 \\
9 $\pm$ 0.500 & 0.431 $\pm$ 0.110 & 2.22 \\
11 $\pm$ 0.500 & 0.654 $\pm$ 0.166 & 2.21 \\
14 $\pm$ 1.00 & 0.0498 $\pm$ 0.0672 & 0.963 \\
17 $\pm$ 2.00 & 0.605 $\pm$ 0.171 & 1.93 \\
24 $\pm$ 4.50 & 0.799 $\pm$ 0.233 & 1.57 \\
33 $\pm$ 3.50 & 0.922 $\pm$  0.579 & 1.57 \\
44 $\pm$ 3.00 & 1.042 $\pm$ 0.817 &  1.61 \\
55 $\pm$  2.00 & 1.643 $\pm$ 0.797 & 1.41 \\
63 $\pm$ 4.00 & 1.97 $\pm$ 1.20 & 1.069 \\
91 $\pm$ 7.50 & 0.738 $\pm$ 1.21 & 0.712 \\
\end{tabular}
\end{center}
\end{table}
	\begin{table}
\caption[Results of weak lensing shear fits.]{Weak lensing shear fit results for high-z ($0.7-1.0$) and all-z ($0.1-1.0$) clusters. This table includes values for richness ($N_{VT}$), mass ($M_{200}$, in units of $10^{14} \; h^{-1} \; M_{\sun}$) and null test (fit to 0) reduced $\chi^2$. Values for concentration ($c_{200}$) are $2.77$ for high-z and $3.01$ for all-z. No corrections for systematics are applied to obtain these values.} \label{allresults2}
\begin{center}
\begin{tabular}{c c c}
{$N_{VT}$}&{$M_{200}$}&$\chi_{red}^2$ \\    
HIGH-Z&&  \\
1 $\pm$ 0 & 0.0200 $\pm$ 0.0375 & 1.01 \\
2 $\pm$ 0 & 0.0519 $\pm$ 0.0505 & 0.931  \\
3 $\pm$ 0 &  0.165 $\pm$ 0.0949 & 1.34   \\
4 $\pm$ 0 & 0.0343 $\pm$ 0.0612 & 1.18   \\
5 $\pm$ 0 & 0.312 $\pm$ 0.173 & 1.91   \\
6 $\pm$ 0 & 0.281 $\pm$  0.196 & 0.774   \\
7 $\pm$ 0 & 0.0533 $\pm$ 0.128 & 0.562   \\
8 $\pm$ 0 & 0.00642 $\pm$ 0.0661 & 1.01   \\
9 $\pm$ 0.500 & 0.680 $\pm$ 0.355 & 0.983  \\
11 $\pm$ 0.500 & 0.258 $\pm$ 0.325 & 0.656   \\
14 $\pm$ 1.00 & 0.212 $\pm$ 0.343 & 1.03   \\
17 $\pm$ 2.00 & 1.56 $\pm$ 0.917 & 0.886   \\
24 $\pm$ 4.50 & 1.66  $\pm$ 1.24 & 1.34   \\
34 $\pm$ 4.50 & 5.31 $\pm$ 4.49 & 1.18   \\
44 $\pm$ 5.00 & 0.0483$\pm$ 1.023 & 1.20   \\ \\
ALL-Z&&  \\
1 $\pm$ 0 & 0.0288 $\pm$ 0.00986 & 1.64   \\
2 $\pm$ 0 & 0.0379 $\pm$ 0.0102 & 1.77   \\
3 $\pm$ 0 & 0.0728 $\pm$ 0.0158 & 2.96  \\
4 $\pm$ 0 & 0.100 $\pm$ 0.0217 & 2.74   \\
5 $\pm$ 0 & 0.144 $\pm$ 0.0282 & 3.20   \\
6 $\pm$ 0 & 0.150 $\pm$ 0.0330 & 2.95   \\
7 $\pm$ 0 & 0.174 $\pm$ 0.0417 & 2.71   \\
8 $\pm$ 0 & 0.212 $\pm$ 0.0502 & 2.58   \\
9 $\pm$ 0.500 & 0.276 $\pm$ 0.0460 & 3.46   \\
11 $\pm$ 0.500 & 0.515 $\pm$ 0.0716 & 5.70   \\
14 $\pm$ 1.00 & 0.304 $\pm$ 0.0618 & 4.052   \\
17 $\pm$ 2.00 & 0.390 $\pm$ 0.0702 & 3.73   \\
24 $\pm$  4.50 & 0.533 $\pm$ 0.0950 & 3.62  \\
35 $\pm$  4.50 &  0.980 $\pm$ 0.202 & 2.78   \\
45 $\pm$  4.50 & 1.21 $\pm$ 0.253 & 2.44   \\
54 $\pm$ 4.50 & 1.93 $\pm$ 0.461 & 2.60   \\
63 $\pm$ 4.00 & 1.82  $\pm$ 0.511 & 2.43   \\
78 $\pm$ 4.00 & 2.12 $\pm$ 0.893 & 1.72   \\
99 $\pm$  4.00 & 0.877 $\pm$ 0.826 & 0.986   \\
\end{tabular}
\end{center}
\end{table}
The value of $N_{VT}$ given is the median of the $N_{VT}$ values in that richness bin. The error on $N_{VT}$ is the range of $N_{VT}$ values, {given by}
\begin{equation}
\small
\Delta N_{VT}=\frac{N_{VT\_max}-N_{VT\_min}}{2}
\end{equation}
The errors on the $M_{200}$ values are standard deviations on values output by the fitting routine.  The $\chi_{red}^2$ we report here is the reduced $\chi^2_{red}$ for the null test of fitting the weak lensing shear to 0. 

	The $\chi^2_{red}$ we report here is not the same as the $\chi_{red}^2$ used for the fit to the NFW profile. What we are checking here is whether the tangential shear profile is well fit by zero shear signal. If this is a good fit (reduced $\chi^2 \approx 1$) then there is no signal. If this is a bad fit (reduced $\chi^2 > 1$), then we have cause to believe there is a tangential shear signal. Note that in the low-z and mid-z bins the $\chi^2_{red}$ are mostly well above 1, which suggests that the fit to 0 is poor and there is measurable weak lensing signal. For the high-z bin many of the $\chi^2_{red}$ values are lower, which suggests the weak lensing signal is less clear for the high-z bin. 
	
\section{Consideration of Systematic Errors}  \label{five}

	We next consider tests to verify our measurements of tangential shear. We also consider systematic errors that can bias measurements of shear and thus measurements of mass and concentration in the NFW model. Our tests to verify our measurements of shear include measurement of orthotangential shear and measurement of tangential shear around random points rather than nearby clusters. Systematics we considered included photo-z bias, the effect of the central BCG, dark matter halo miscentering, uncertainties on photo-zs and the contamination of source galaxies by misidentified lensing galaxies.
	
\subsection{Orthotangential Shear Test}
	For each richness and redshift bin we made plots of both tangential and orthotangential shear. For tangential shear we expect to see an increase in shear at small distances from the cluster center (consistent with non-zero tangential shear) while for orthotangential shear we expect to see no change in the orthotangential shear at small distances (consistent with zero orthotangential shear). If we see significant signal in the orthotangential shear, this can be evidence of error as weak lensing will not produce orthotangential shear. As seen in Figure \ref{sampleplot} typically we can see a clear signal in tangential shear while the orthotangential shear is apparently consistent with zero.
	
\begin{figure}   
\begin{center}
\includegraphics[scale=0.37, angle=90]{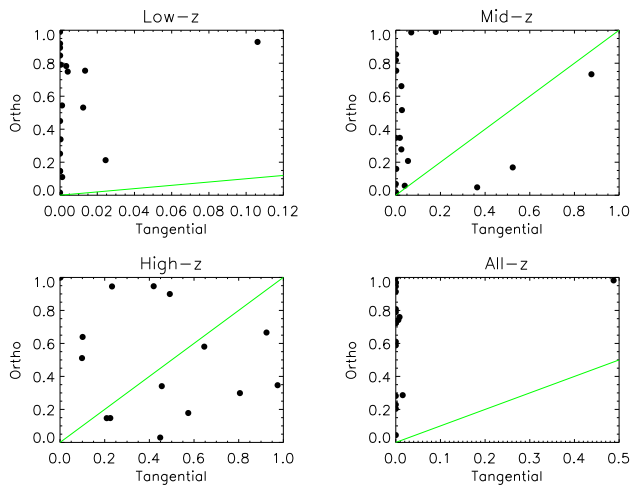}
\caption[$\chi^2$ probabilities compared for tangential and orthotangential shear.]{$\chi^2$ probabilities compared for tangential and orthotangential shear. The green line is the $y=x$ line.} \label{tanortho}
\end{center}
\end{figure}

	In Figure \ref{tanortho} we compare the $\chi^2$ probability calculated for the tangential shear to that calculated for orthotangential shear. 
This quantity tells us the probability that the $\chi^2_{red}$ for a fit to 0 will exceed the given value of $\chi^2_{red}$. For null data, like we expect orthotangential shear signal to be, this probability should be on the order of 0.5, since we expect the orthotangential shear to be well fit to 0. For tangential shear this probability should be very small, since we would expect tangential shear measurements to not be consistent with 0. Note that in Figure \ref{tanortho} we find that $\chi^2$ probability is almost always significantly higher for orthotangential shear than it is for tangential shear. For orthotangential shear, the probability ranges from 0 to 1, and so on average is about 0.5, as expected. In most cases, the probabilities for tangential shear are near 0, meaning they are not well fit to a tangential shear of 0. Thus we conclude that in most cases, there is evidence for tangential shear signal, and not for orthotangential shear signal. The one notable exception is the high-z ($0.7 \leq z \leq 1.0$) data. The high-z data has the lowest statistics of any of the bins, and so the tangential shear fits have higher error bars. The tangential shear signal is less obvious, with $\chi^2$ probabilities from 0 to 1, while the orthotangential shear signal is unchanged. 

\subsection{Random Points Test}
	We would expect no tangential shear signal around random points in the sky while we would expect a non-zero tangential shear around clusters. If random points were to give signal similar to that measured near clusters, this would be evidence that there is a systematic problem in our measurements. In order to conduct this random points test we first generated a sample of random points. {We generated 19 706 random values} for RA and DEC within the ranges of the real values, with a distribution of $N_{VT}$ and $z$ similar to that found in the real data.

\begin{figure}  
 \centering
 \textbf{REAL POINTS}  \\
\begin{subfigure}
  \centering
\includegraphics[scale=0.37, angle=90]{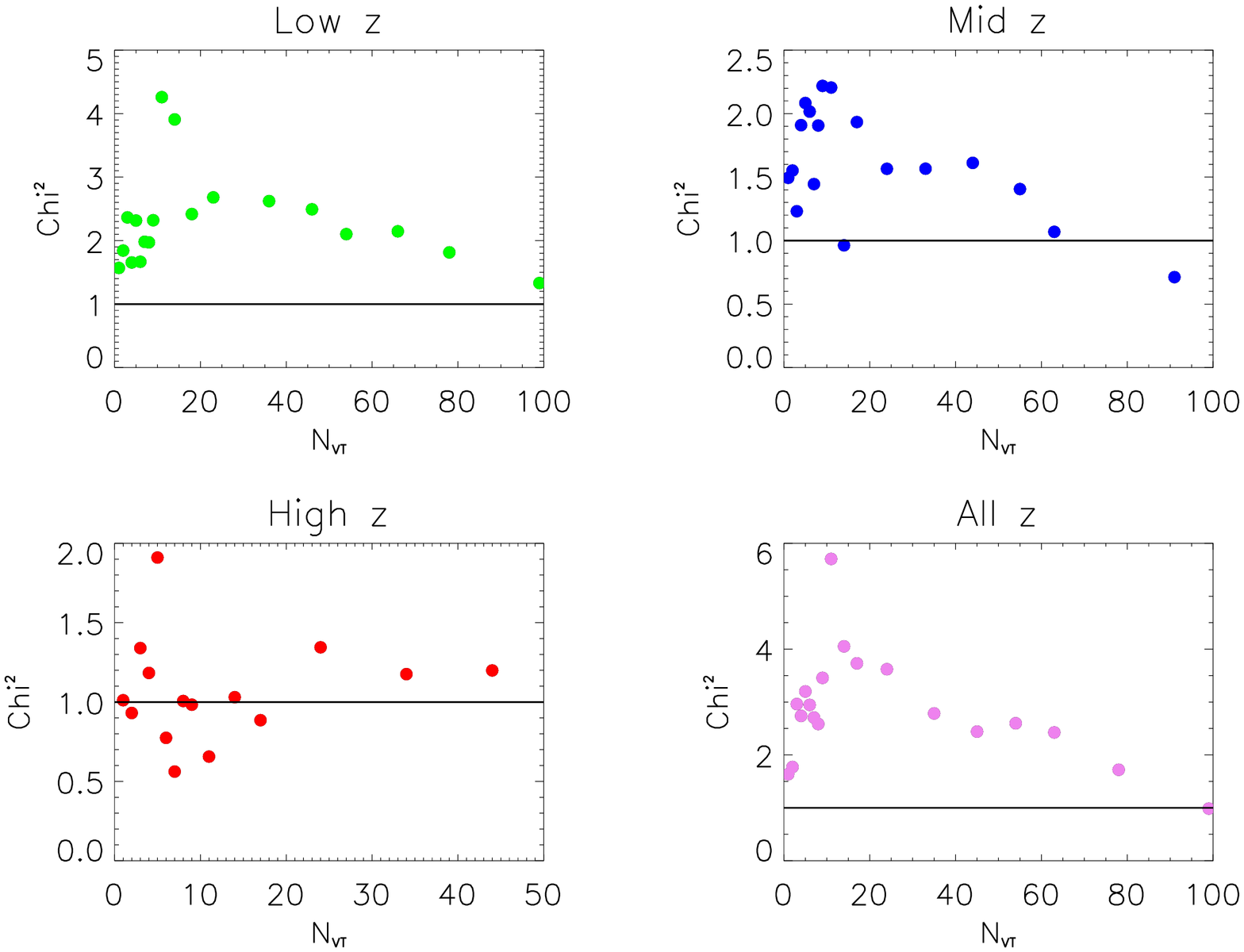}
\end{subfigure}
\\
\centering
\begin{subfigure} 
\centering
\textbf{RANDOM POINTS}  \\
\includegraphics[scale=0.37, angle=90]{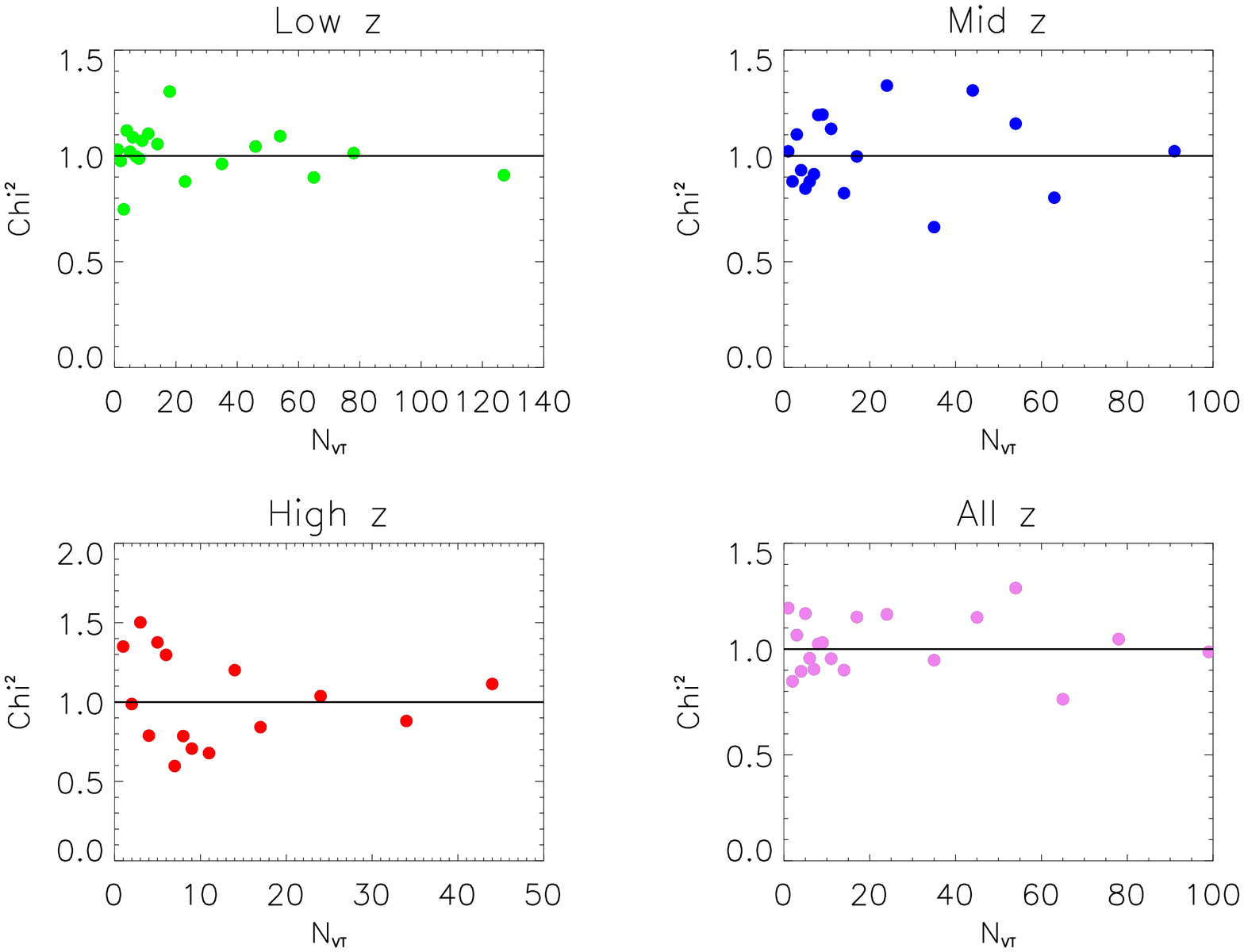}
\end{subfigure}
\caption[Reduced $\chi^2$ for real and random data.]{Reduced $\chi^2$ values for tangential shear fits to zero for clusters and for the same fits to random points, given compared to $N_{VT}$. Note that for the clusters, the reduced $\chi^2$ values are well above 1, meaning they are not consistent with a model of no shear signal. For the random points, reduced $\chi^2$ values are close to 1, suggesting no evidence of shear signal (as expected).} \label{chicompare} 
\end{figure}

	Once we generated the random points catalog, we remeasured tangential shear using the random points as the clusters. We found that the tangential shear profiles generated were now consistent with zero, as would be expected for random points. In Figure \ref{chicompare} we plot the reduced $\chi^2$ values from tangential shear fits to zero for both the actual data (upper) and for the random points (lower). 
Note that for the real data the reduced $\chi^2$ values are mostly larger than 1, meaning that the real data is not consistent with fit to zero, that is a model of no shear signal. On the other hand, for the random points the reduced $\chi^2$ values are mostly near 1, meaning that the random points are consistent with no tangential shear. This reinforces our claim that we have measured a significant weak lensing shear signal. 
	
	\subsection{Photo-z Bias}
	A photometric redshift (or photo-z) is a redshift measured by observing magnitudes in multiple filters and then using a neural network or other method to estimate redshift based on this information and a training set of spectroscopic redshifts (spec-z). The relation between the magnitudes measured in different filters has a dependence on the redshift of an object, thus photometry can be used to estimate redshifts. Photometric redshifts are usually used in large surveys where it is implausible to find spectroscopic redshifts for most objects. The advantage to this method is that redshifts can be estimated for many objects, but the disadvantage is that errors on photometric redshifts are significantly larger than those for spectroscopic redshifts. We consider two systematic effects caused by photo-zs. In this section we consider a small bias in the photo-zs and in Section \ref{findsigmacrit} we consider the effect of photo-z uncertainties on the value of $\Sigma^{-1}_{crit}$.   
	
\begin{figure}   
\begin{center}
\includegraphics[scale=0.4, angle=0]{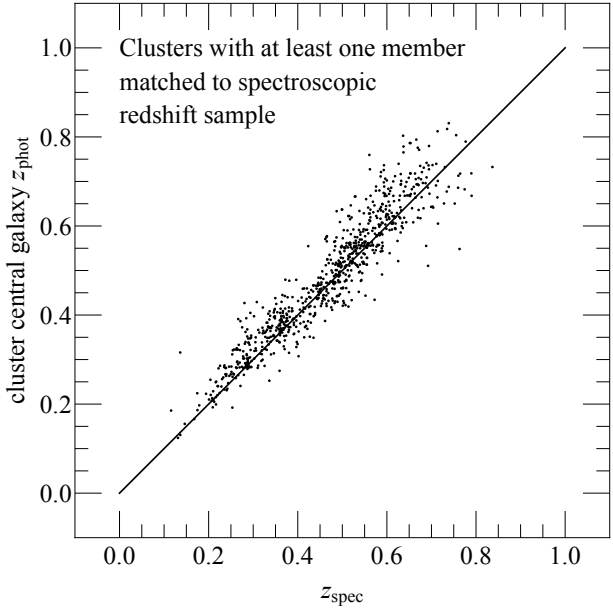}
\caption{A comparison of photometric redshift to spectroscopic redshift for clusters that had at least one member with a spectroscopic redshift. The photo-zs were the redshift of the central BCG in each cluster. This relation was used to find photo-z bias corrections for each redshift bin.}  \label{marcelle}
\end{center}
\end{figure}

	When comparing the photometric redshifts of VT clusters to maxBCG cluster redshifts and spectroscopic redshifts for some cluster galaxies we found a small bias (see Figure \ref{marcelle}). We found that the cluster photometric redshifts were too high by 0.0277 in the low-z bin, by 0.0348 in the mid-z bin, by 0.00974 in the high-z bin and by 0.0308 in the all-z bin. In order to correct for this, we subtracted these values from all cluster redshifts before these redshifts were used for binning and for calculation of $\Sigma^{-1}_{crit}$. 

	\subsection{Central BCG}  \label{centralbcg}
\begin{figure}   
\begin{center}
\includegraphics[scale=0.37, angle=90]{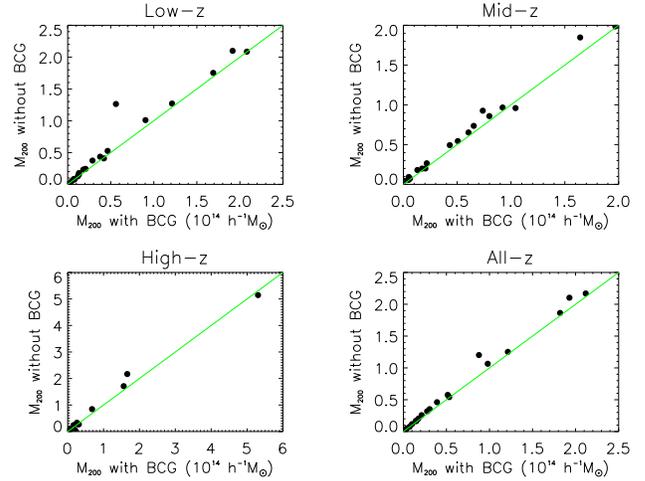}
\caption[$M_{200}$ with and without the central $0.1 \; h^{-1}$ Mpc included in fitting.]{$M_{200}$ with and without the central $0.1 \; h^{-1}$ Mpc included in fitting. The green line is the $y=x$ line. Note that the points mostly follow this line.}  \label{missingBCG}
\end{center}
\end{figure}

	In our fitting of the mass and concentration, we fit an NFW profile to the dark matter halo of the clusters. In so doing, we did not include a term for the mass of the central BCG. Thus we wanted to try to estimate what effect the mass of the BCG would have on the shear profile. To do this we excluded the central $0.1 \; h^{-1}$ Mpc. We did this as the mass of the BCG would usually affect only the innermost regions of the shear profile. In Figure \ref{missingBCG} we compare the cluster masses found with and without the central $0.1 \; h^{-1}$ Mpc region of the clusters. 
The green line is the $y=x$ line. The points follow this line quite closely, with a few exceptions. Thus we conclude that the mass of the central BCG has minimal impact on the shear profiles. Nevertheless for our final measurements we exclude the central $0.1 \; h^{-1}$ Mpc of the stacked cluster profiles from the NFW fits in order to avoid bias from the mass of the central BCG.
	
	\subsection{Halo Miscentering}  \label{halomiscenter}
	When we fit an NFW profile to the dark matter halos of the galaxy clusters, we take the BCG to be the center of the halo, measuring all distances with respect to this. However, as mentioned in Section \ref{one}, this will be incorrect some fraction of the time, as the BCG at times is not the center of the dark matter halo. If mass is measured at a location outside the center of a halo, the mass found for that halo will be less than it should be, since the outer region of a halo is then being taken as the center. As this will bias mass measurements, our analysis needs to include a consideration of the contribution from halos that are not centered on the BCG. In order to consider halo miscentering we use Equation \ref{simet_miscenter} \citep{Simet11}. 
	
	{\citet{Simet11} base their method} on the maxBCG method, and so their probability distribution of miscentering in dark matter halos will follow the distributions discussed in \citet{Johnston07}. We examined whether the probability of miscentering of clusters in halos would be the same for our VT method as it was for the maxBCG method. We found a very similar miscentering distribution using simulations described and used in \citet{Soares10}, with a 2D Gaussian fit giving a standard deviation $\sigma=0.47 \; h^{-1}$ Mpc while \citet{Johnston07} found $\sigma=0.42 \; h^{-1}$ Mpc.   
	
\begin{figure}   
\begin{center}
\includegraphics[scale=0.37, angle=90]{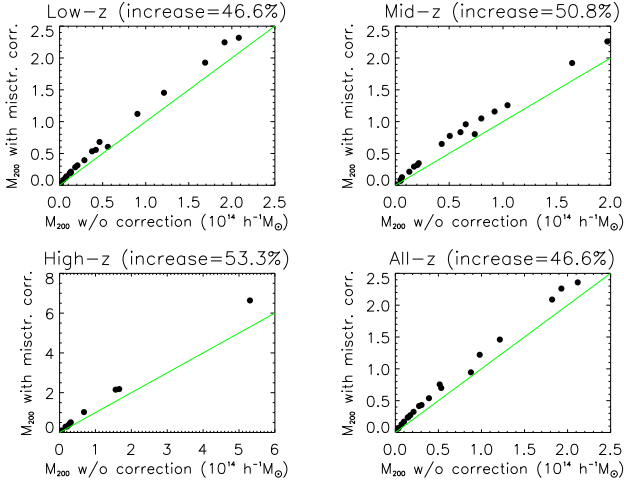}
\caption[$M_{200}$ values before and after miscentering correction.]{$M_{200}$ values before and after miscentering correction. The green line is the $y=x$ line. Note that all masses are pushed upward by about $50\%$.} \label{missCENTER}
\end{center}
\end{figure}

	We applied Equation \ref{simet_miscenter} by taking the masses output by the fitting routine as $M_{200,mis}$ and multiplying them by $1.44$. Then the true masses ($M_{200,true}$) are a function of $N_{200}$, following Equation \ref{simet_miscenter}. Before we calculated $M_{200,true}$ we had to convert $N_{VT}$ to $N_{200}$ using Equation \ref{convertNVT}. In Figure \ref{missCENTER} we compare masses found without the miscentering correction to masses found after the miscentering correction. 
Note that all masses are systematically higher. For each of the redshift bins we found the ratio of miscentering-corrected mass to the uncorrected masses and then found the median of the ratios. This median is shown in Figure \ref{missCENTER} as the quantity \textit{increase}. Note that the percent increases are signficant and range from $46.6\%$ to $53.3\%$. 
	
\subsection{Uncertainties on Photometric Redshifts}  \label{findsigmacrit}
	All source galaxies in the sample had photometric redshifts measured for them. Our weak lensing mass measurements depend on photometric redshifts because tangential shear $\gamma$ depends on $\Sigma^{-1}_{crit}$, which is a function of redshift. Thus we need to consider the errors which may be introduced into our measurements by uncertainties on photometric redshifts. To do this, we assembled a sample of spectroscopic redshifts to compare to.
	
	We gathered public data from four data sets:  The Stripe 82 co-add spectroscopic data, VVDS, DEEP2 and VIPERS. To access spectroscopy from the Stripe 82 co-add we used the SDSS-III catalog archive server to query for spectroscopic redshifts for objects within Stripe 82 that had i-band magnitudes $19 \leq i \leq 23.5$, type=3 (galaxies) and zwarning=0. The last parameter is a measure of the quality of the spectroscopic redshift. DEEP2 \citep{DEEP2} is a project which uses the Keck telescopes at Mauna Kea. The DEEP2 project seeks to understand the evolution of galaxies and galaxy clusters by taking spectra of more than 50 000 galaxies out to redshift greater than 0.7. The VIMOS VLT Deep Survey (VVDS) \citep{VVDS} is a survey conducted using the VIMOS (VIsible imaging Multi-Object Spectrograph) instrument on the Very Large Telescope (VLT). VVDS observed 3 fields that observed 35 526 galaxies of magnitude ranging between 17.5 and 24.75 and redshift between 0 and 6.7. The VIMOS Public Extragalactic Redshift Survey (VIPERS) \citep{VIPERS} is a survey of high-redshift objects taken using the VIMOS spectrograph on the VLT. VIPERS observed two fields of objects chosen from the Canada-France-Hawaii Telescope Legacy Survey-Wide (CFHTLS-Wide) optical photometric catalog. In the first public data release of VIPERS there are 54 756 galaxies with spectroscopy. In each of these surveys we match the galaxy catalog with our own catalog of galaxies in Stripe 82. Our final catalog of objects with spectroscopic redshifts is a list of all objects from these 4 sets of data that overlap with objects in the Stripe 82 co-add. We found a total of 34 033 matched source galaxies with spectroscopic redshifts.   	 

	In order to better constrain the effect of photo-z error, we analyzed how $\Sigma_{crit}^{-1}$ changed with different measures of redshift. We looked at four different ways of finding $\Sigma_{crit}^{-1}$ and then we compared them to see how much $\Sigma_{crit}^{-1}$ varied based on the method:
\begin{enumerate}
\item Find $\Sigma_{crit}^{-1}$ using the median photo-z of the source galaxies.
\item Find $\Sigma_{crit}^{-1}$ for each source galaxy and then take the average, using photo-zs.
\item Find $\Sigma_{crit}^{-1}$ for each galaxy that has a spec-z but use the photo-z for each of those galaxies. Photo-zs must be weighted {to match the distribution} of the much larger Stripe 82 photo-z sample (see details below).
\item Find $\Sigma_{crit}^{-1}$ for each galaxy that has a spec-z but use the spec-z for each of those galaxies. Spec-zs must also be weighted, as in (iii).
\end{enumerate}

	We would expect method (iv) to give the best results since it uses spectroscopic redshifts, but since the number of spectroscopic redshifts is small, we have to first demonstrate that the weighting method works. In order to demonstrate this, we use methods (ii) and (iii), which both use photo-zs. Method (ii) involves finding $\Sigma_{crit}^{-1}$ for each source galaxy and method (iii) involves finding it for galaxies that have spectroscopic redshifts but with weighting. If they match, then this suggests that the weighting works well, and can then be used with the spectroscopic redshift sample.

	The original method (method (i)) by which we found $\Sigma^{-1}_{crit}$ was to find the photometric redshift of the cluster and then find the median of all of the photometric redshifts of the source galaxies. These redshifts were used to find the angular diameter distances, then $\Sigma^{-1}_{crit}$ was calculated from them. Galaxies were considered source galaxies if they were (1) within $3 \; h^{-1}$ Mpc of the BCG, (2) had i-band magnitude between 20.5 and 23.5, (3) had ellipticities less than 2 and (4) had $z_{source} > z_{cluster}+0.1$. To test this method we produced a table of 1000 redshifts from $0.1-1.0$, treating these as cluster redshifts. We then took the first 1000 objects in the source galaxy catalog and looped through these to find source galaxies to correspond to each of the fake clusters; we only required that the source galaxies have $20.5 \leq i \leq 23.5$ and that $z_{source} > z_{cluster}+0.1$. We then took the median photo-z of these source galaxies and used this number along with redshift of the cluster to find $\Sigma^{-1}_{crit}$. 

	For method (ii) (using all photometric redshifts) we used almost the same method as in the previous Section. However rather than finding the median source redshift, for each cluster we looped through all source galaxies and found $\Sigma^{-1}_{crit}$ for each. Then for each cluster we found a mean value of $\Sigma^{-1}_{crit}$.

	In method (iii) we began to use the 34 033 galaxies for which we had both photometric and spectroscopic redshifts. However since we were using this small spectroscopic sample to find values of $\Sigma^{-1}_{crit}$ for the full Stripe 82 co-add dataset, we had to weight the small spectroscopic sample to match the distribution of the much larger Stripe 82 sample. We utilized a routine in C++ called $\tt CALCWEIGHTS.CPP$ developed by C. Cunha that uses a nearest-neighbor code to extrapolate a redshift distribution \citep{Lima08}.
  
	Now that we had these weights we could find $\Sigma^{-1}_{crit}$ using method (iii). We produced a cluster sample like before, ranging in redshift from $0.1-1.0$. The source redshifts we used in method (iii) were the photometric redshifts that corresponded to the 34 033 galaxies with spectroscopic redshifts. Then for each cluster redshift we looped through all the (photometric) source redshifts, producing a value of $\Sigma_{crit}\times weight$ for each.  Finally we found the mean by finding for $N$ source galaxies: 
\begin{equation}  \label{meanie}
\Sigma^{-1}_{crit\_ave}=\left(\frac{\sum\limits_{i=1}^N \Sigma_{crit\_i} \times weight_i}{\sum\limits_{i=1}^N weight_i}\right)^{-1}
\end{equation}          

	Finally in method (iv), we used the same method as method (iii), but this time the source redshifts were the spectroscopic redshifts. Again we found $\Sigma_{crit}\times weight$ for each source galaxy for a particular cluster redshift and again we found the mean value of $\Sigma^{-1}_{crit}$ for that cluster redshift by using Equation \ref{meanie}.

\begin{figure}   
\begin{center}
\includegraphics[scale=0.37, angle=90]{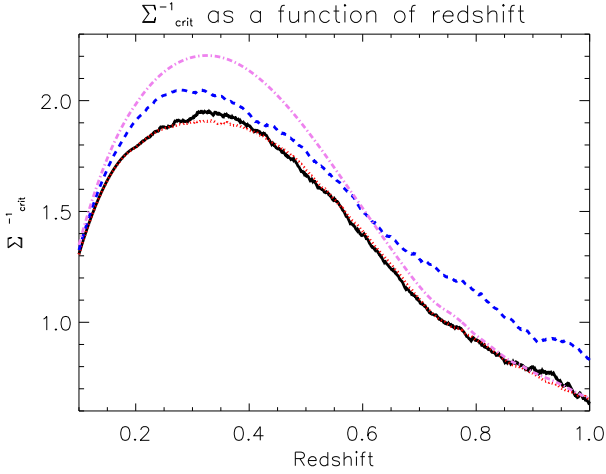}
\caption[$\Sigma_{crit}^{-1}$ as a function of redshift.]{$\Sigma_{crit}^{-1}$ as a function of redshift for each of the four different methods described in this section. The dash-dot violet line is from method (i), the solid black line is from method (ii), the dotted red line is from method (iii) and the dashed blue line is from method (iv). All values of $\Sigma_{crit}^{-1}$ are multiplied by $10^{-16}$ $m^2/kg$.} \label{sigmacritplot}
\end{center}
\end{figure}

	Our final product in each case was a lookup table of values of $\Sigma^{-1}_{crit}$ for a set of cluster redshifts ranging from $0.1-1.0$. In Figure \ref{sigmacritplot} we plot $\Sigma_{crit}^{-1}$ as a function of redshift from $0.1-1.0$. 
In this plot the dash-dot violet line is from method (i), the solid black line is from method (ii), the dotted red line is from method (iii) and the dashed blue line is from method (iv). Note that the original method of finding $\Sigma_{crit}^{-1}$ (method (i)) deviates the most from the other methods. The red and the black lines ((methods (ii) and (iii)) match closely, indicating that the weighting method effectively reproduces the larger sample.	

	We finally conclude that there are significant differences in $\Sigma_{crit}^{-1}$ depending on which method of finding it we use. We also conclude, from the closeness of the black and the red lines, that the weights code is highly effective at reproducing the photometric distribution from the spectroscopic distribution. Therefore we finally chose to use the weighted distribution of spectroscopic redshifts (the blue line from method (iv) in Figure \ref{sigmacritplot}) to calculate $\Sigma_{crit}^{-1}$.
	
	We produced a lookup table of values of $\Sigma^{-1}_{crit}$ using these spectroscopic redshifts and then had the fitting routine refer to this table when finding $\Sigma^{-1}_{crit}$ for a particular cluster. The scale factor (Equation \ref{scaleslifted}) was then calculated as:
\begin{equation}
S=\frac{\Sigma_{crit\_actual}}{\Sigma_{crit\_fiducial}}
\end{equation}  
$\Sigma_{crit\_fiducial}$ was taken as $7.94\times10^{15}$ $m^2kg^{-1}$, which is the value of $\Sigma_{crit}$ found for $z_{source}=0.75$ and $z_{cluster}=0.55$.	 

\subsection{Foreground Galaxy Contamination} \label{contaminated}

	Weak lensing shear was measured for a sample of galaxies that were more distant than each of the lensing clusters. We measured shear only for galaxies that were at a redshift that was 0.1 greater than cluster redshift. The redshifts used for these cuts are photometric redshifts, and thus they have larger errors than if they were spectroscopic redshifts. These errors can allow some galaxies that are actually at lower redshift than the cluster (i.e., in front of the cluster) to be included in the sample of background source galaxies. This would contaminate the shear signal with galaxies that cannot exhibit shear. This is especially a problem nearby a galaxy cluster, as cluster galaxies can be misidentified as background galaxies. The contamination can be quantified by noting that average tangential shear ($\gamma_{ave}$) in a particular richness bin is
\begin{equation}
\gamma_{ave}=\frac{\sum\limits_{i=1}^{N_{real}} \gamma_{T\_real}+\sum\limits_{i=1}^{N_{fake}} \gamma_{T\_fake}}{N_{real}+N_{fake}}
\end{equation}  
where \textit{real} and \textit{fake} refer to source galaxies and misidentified foreground galaxies respectively and $N$ means total number of galaxies found in that richness bin. We expect $\Sigma \gamma_{T\_fake}$ to sum to zero since this includes only shape noise and no actual shear. However, the $N_{fake}$ term remains, and we need to find a correction factor $C(r)$ to account for it. 

	What we want is to remove the $N_{fake}$ term in the denominator. Thus we need to multiply the measured $\gamma_{ave}$ by this term:
\begin{equation}
C(r)=\frac{N_{real}+N_{fake}}{N_{real}}
\end{equation}
But since we cannot directly measure $N_{real}$, we can instead measure a number density, $n$, the number of galaxies per area per cluster. Thus
\begin{equation} \label{foregroundcorrect}
C(r)=\frac{N_{real}+N_{fake}}{N_{real}}=\frac{n_{measured}(r)}{n_{random}}
\end{equation}
Here $n_{measured}(r)$ is the number of galaxies measured per area per cluster at various radii around clusters; this tells us about $N_{real}+N_{fake}$ because it accounts for all apparently source galaxies found nearby clusters (including misidentified cluster galaxies). The second term, $n_{random}$, is the number of galaxies per area per cluster measured at random points; that is, far enough away from the center of the cluster that the concentration of galaxies that marks a galaxy cluster does not affect the number density. 

	Number density $n$ is measured by finding the total number of galaxies per area per cluster in a given richness bin. To find this, we find the total number of source galaxies in bins of $0.1 \; h^{-1}$ Mpc of distance from the BCG in each richness and redshift bin. We then divide this number by the product of the area of an annulus at that increment of distance from the BCG and the number of clusters in that richness bin. In other words, we measure
\begin{equation}
n=\frac{\frac{N_{total}}{A_{annulus}}}{N_{clusters}}
\end{equation}	
where $N_{total}$ is the total number of apparently source galaxies in that richness and redshift bin, $A_{annulus}$ is the area of the annulus and $N_{clusters}$ is the number of clusters in that bin. 

	When we first measured the number densities, we found that they all decreased with radius, even for the random points. This is unexpected behavior, as random points should overall have no dependence of number density on position. It was found that this behavior was due to the nature of Stripe 82 as a thin strip of observations (2.5 degrees wide in declination). For clusters that were near the edge of the images the number of nearby galaxies that could be measured would be limited to those that appeared on the images. As the distance from the cluster center increased, the number of galaxies per area would drop, not due to an actual decrease but simply because we had progressed outside the available imaging region. To address this issue, we remeasured the number of source galaxies per area per cluster while rejecting any clusters for which a circle of radius $3.0 \; h^{-1}$ Mpc would be off the imaging area. 

\begin{figure}  %Made by /data/des30.a/data/mwiesner/RERUN/CORRECTIONS/plot_corrections.pro
  \centering
\includegraphics[scale=0.35, angle=90]{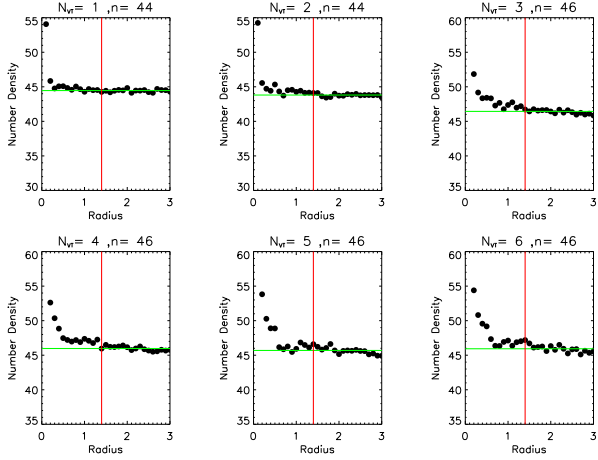}
\caption{Plots showing number density as a function of radius found for the first six richness bins for $0.4\leq z \leq 0.7$. The horizontal green line represents the median value of number density and the vertical red line represents the beginning radius at which the random number density was measured. \label{faraway}} 
\end{figure}

	We found that number density for random points in the sky is about the same as the number density in the region of a cluster but far away from the center. Because of this we chose to use the number density in the region of clusters but far away as the reference point for our foreground corrections. As the number density approached a constant by $1.4 \; h^{-1}$ Mpc, we measured the random number density between $1.4-3.0 \; h^{-1}$ Mpc (see Figure \ref{faraway}).

	We finally multiplied the average tangential shear ($\gamma_{ave}$) by the $C(r)$ at each radius. The values of $C(r)$ usually ranged from about {$1.5$ at the most to $0.9$} at the least. The corrections are applied for each radius bin (of size $0.1 \; h^{-1}$ Mpc), but beyond the first several radius bins most corrections are close to 1 (meaning no change). 
\\
\section{Mass-Richness Relations}  \label{six}
	\subsection{Measuring a Mass-Richness Relation}
		In order to find a mass-richness relation by using the weak lensing shear results, we fit a power relation to $M_{200}$ and $N_{VT}$ values. In each case we found mass-richness relations for many richness bins in four separate cluster redshift bins, $0.1-0.4$ (low-z), $0.4-0.7$ (mid-z), $0.7-1.0$ (high-z) and $0.1-1.0$ (all-z). Note that as mentioned in Section \ref{one}, all mass-richness relations are given in the form:
\begin{equation}
M_{200}=M_{200|20}\left( \frac{N_{VT}}{20}  \right)^{\alpha}
\end{equation}   
where $\alpha$ is the slope of the power law. This parameter tells us how quickly mass increases with an increase in richness. $M_{200|20}$ is the mass coefficient; we define this to be the value of $M_{200}$ when $N_{VT}=20$. 
	
\begin{figure}
\begin{center}
\includegraphics[scale=0.35, angle=90]{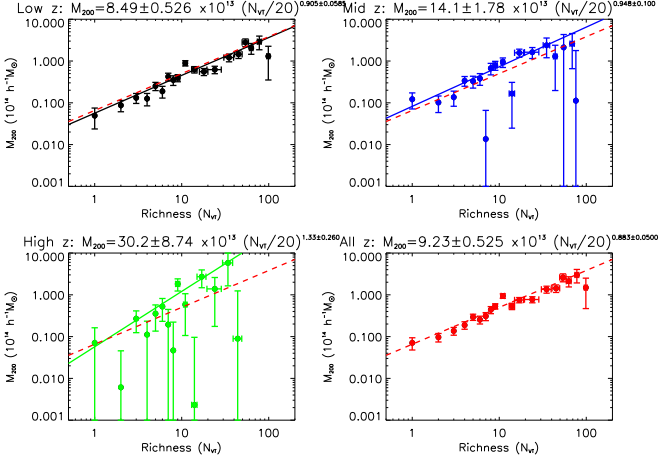}
\caption[Plot of $M_{200}$ vs. $N_{VT}$ including corrections for all systematics.]{The mass-richness relation ($M_{200}$ vs. $N_{VT}$) for low-z (upper left), mid-z (upper right), high z (lower left) and all-z (lower right), including corrections for photo-z bias, the effect of the central BCG, halo miscentering, photo-z uncertainty and foreground galaxy contamination. The dotted red line is the fit for all redshift bins (all-z) and is included for comparison in each case.} \label{mass_rich_6} 
\end{center}
\end{figure}

	Our initial mass-richness relations for the Stripe 82 co-add were found with no corrections for the systematics discussed in Section \ref{five}. We then also found the mass-richness relations for the Stripe 82 co-add including corrections for all of the above systematics. We corrected for the bias in the photometric redshifts. We excluded the central $0.1 \; h^{-1}$ Mpc in the fits, using only source galaxies between $0.1-3.0 \; h^{-1}$ Mpc from the BCG. We included the correction for halo miscentering. We also used the spectroscopic redshift lookup table to find values for $\Sigma_{crit}$ and we applied the foreground galaxy correction factors when calculating $\gamma_{ave}$. Our final plot of mass vs. richness is shown in Figure \ref{mass_rich_6} and the final numerical values for the mass-richness relations are shown in Table \ref{massrich1}. We note that the errors are smallest for the low-z and all-z bins, while they are largest for the high-z bin. 
\begin{table} 
\setlength{\tabcolsep}{1pt} 
\caption[Mass-richness relations for Stripe 82.]{Mass-richness {relation fit parameters} for the Stripe 82 co-add. Mass coefficients all have the units $10^{13} \; h^{-1} \; M_{\sun}$. The $\chi^2_{red}$ describes the goodness of the power law fit to the mass-richness data.} \label{massrich1}
\begin{center}
\begin{tabular}{c c c c}
\head{2.0cm}{Redshift Bin}&\head{2.0cm}{Mass Coefficient}&\head{2.0cm}{Slope of Power Law}&\head{2.0cm}{$\chi^2_{red}$} \\  
No corrections  &&& \\

0.1--0.4&5.23 $\pm$ 0.375& 1.05 $\pm$ 0.0632&1.10 \\  %UPD
0.4--0.7&7.46 $\pm$ 0.917&0.910 $\pm$ 0.0996&1.28  \\ 
0.7--1.0&16.9 $\pm$ 5.75&1.38 $\pm$ 0.271&0.351 \\
0.1--1.0&5.78 $\pm$ 0.368&1.04 $\pm$ 0.0548&1.22 \\

Photo-z bias  &&& \\

0.1--0.4&4.71 $\pm$ 0.322& 1.02 $\pm$ 0.0598&1.52 \\  %UPD
0.4--0.7&7.48 $\pm$ 1.06&1.10 $\pm$ 0.118&0.728  \\ 
0.7--1.0&20.0 $\pm$ 6.63&1.45 $\pm$ 0.280&0.548 \\
0.1--1.0&5.10 $\pm$ 0.322&1.04 $\pm$ 0.0528&1.45 \\

Central BCG &&& \\
 
0.1--0.4&5.85 $\pm$ 0.402& 1.03 $\pm$ 0.0642&1.18 \\ % UPD
0.4--0.7&8.17 $\pm$ 0.969&0.899 $\pm$ 0.0931&1.32  \\ 
0.7--1.0&18.8 $\pm$ 6.10&1.31 $\pm$ 0.259&0.425 \\
0.1--1.0&6.39 $\pm$ 0.394&1.02 $\pm$ 0.0539&1.31 \\ 

Halo Miscentering &&& \\

0.1--0.4&7.02 $\pm$ 0.495& 0.904 $\pm$ 0.0628&1.13  \\  %UPD
0.4--0.7&10.0 $\pm$ 1.22&0.768 $\pm$ 0.0990&1.28 \\ 
0.7--1.0&22.8 $\pm$ 7.73&1.25 $\pm$ 0.271&0.352 \\
0.1--1.0&7.76 $\pm$ 0.487&0.901 $\pm$ 0.0538&1.26\\
  
Photo-z uncertainty &&& \\
 
0.1--0.4&5.73 $\pm$ 0.410& 1.06 $\pm$ 0.0641&1.10 \\   % UPD
0.4--0.7&8.14 $\pm$ 0.999 &0.933 $\pm$ 0.0998&1.27 \\ 
0.7--1.0&12.9 $\pm$ 4.39&1.43 $\pm$ 0.281&0.347  \\
0.1--1.0&6.26 $\pm$ 0.401&1.07 $\pm$ 0.0547&1.20  \\
 
Foreground galaxies &&& \\
 
0.1--0.4&5.66 $\pm$ 0.391& 1.08 $\pm$ 0.0612&1.15 \\    %UPD
0.4--0.7&8.22 $\pm$ 0.961&0.897 $\pm$ 0.0948&1.20 \\ 
0.7--1.0&19.6 $\pm$ 6.10&1.33 $\pm$ 0.240&0.529 \\
0.1--1.0&6.28 $\pm$ 0.388&1.04 $\pm$ 0.0511&1.33\\
 
All corrections  &&& \\
 
0.1--0.4&8.49 $\pm$ 0.526&0.905 $\pm$ 0.0585&1.89 \\   %UPD
0.4--0.7&14.1 $\pm$ 1.78&0.948 $\pm$ 0.100&0.868 \\
0.7--1.0&30.2 $\pm$ 8.74&1.33 $\pm$ 0.260&0.674\\
0.1--1.0&9.23 $\pm$ 0.525&0.883 $\pm$ 0.0500&1.94 \\
     \end{tabular}
\end{center}
\end{table}	

Finally we present all mass-richness relations found including all systematics for (1) low-z, (2) mid-z, (3) high-z and (4) all-z:
\begin{equation}
\begin{array}{l}
(1) \;\; M_{200}=(8.49 \pm 0.526)\times10^{13} \; h^{-1} M_{\sun}\left(\frac{N_{VT}}{20}\right)^{0.905\pm 0.0585}  \\
(2) \;\; M_{200}=(14.1 \pm 1.78)\times10^{13} \; h^{-1} M_{\sun}\left(\frac{N_{VT}}{20}\right)^{0.948\pm 0.100}  \\
(3) \;\; M_{200}=(30.2 \pm 8.74)\times10^{13} \; h^{-1} M_{\sun}\left(\frac{N_{VT}}{20}\right)^{1.33\pm 0.260}  \\
(4) \;\; M_{200}=(9.23 \pm 0.525)\times10^{13} \; h^{-1} M_{\sun}\left(\frac{N_{VT}}{20}\right)^{0.883\pm 0.0500}  \\
\end{array}
\end{equation}  

	\subsection{Comparison of Results}
	\subsubsection{The Effects of the Systematics}
\begin{figure}  
%Made by compare_mr.pro
\begin{subfigure}
MASS COEFFICIENT \\
\centering
\includegraphics[scale=0.35, angle=90]{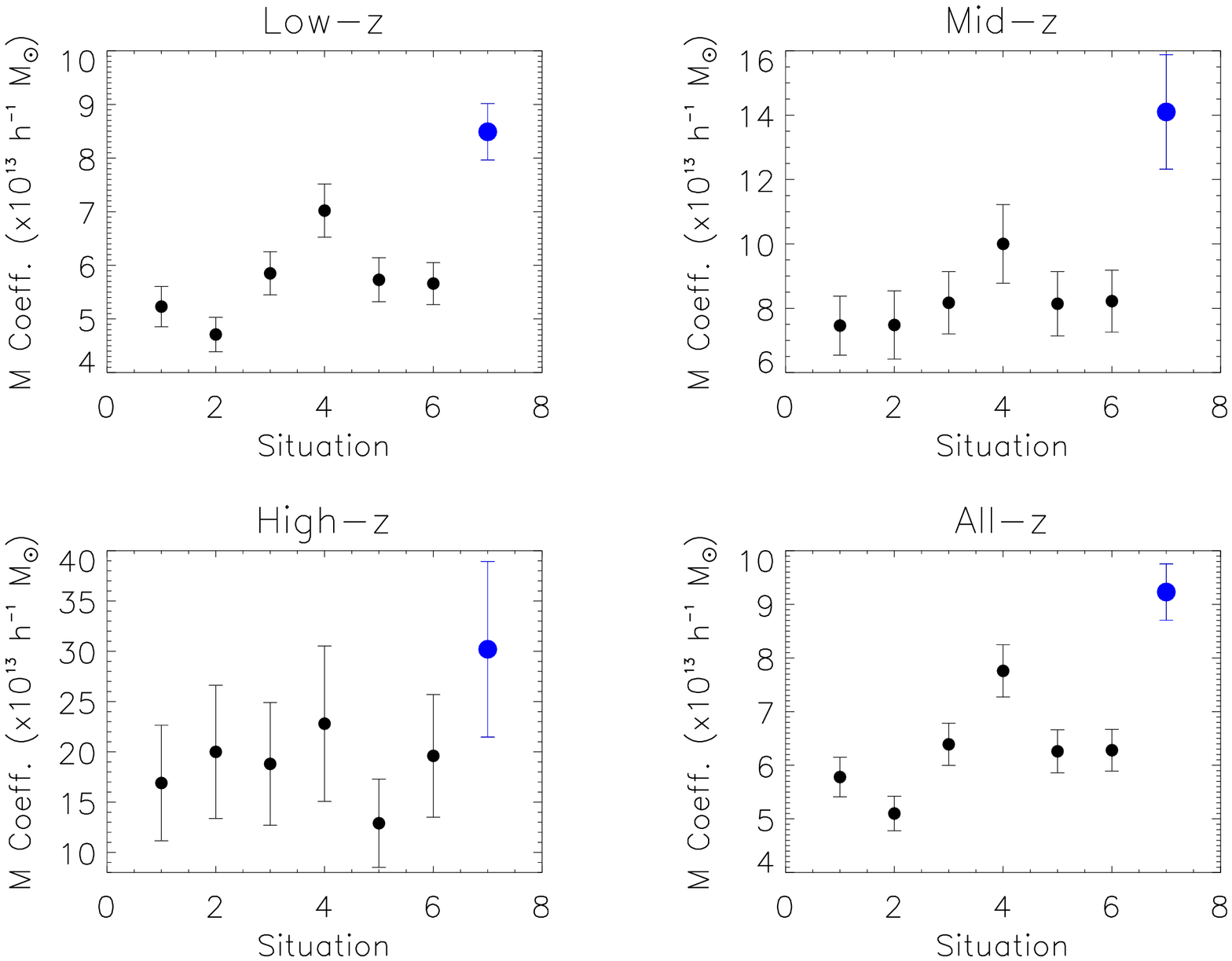}
\end{subfigure} \\
\begin{subfigure}
POWER LAW SLOPE \\
\centering
\includegraphics[scale=0.35, angle=90]{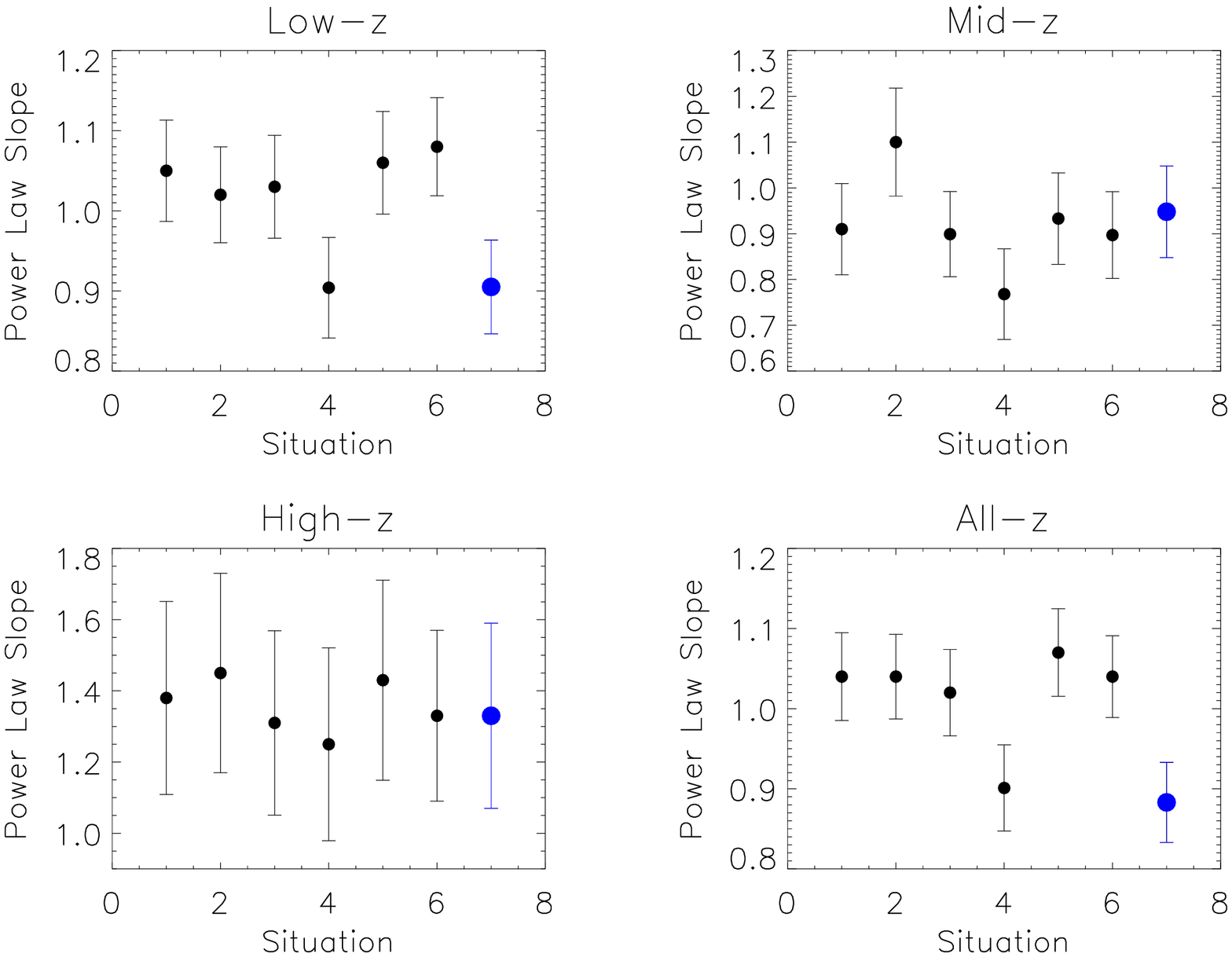}
\end{subfigure}\\
\caption[Comparison of Stripe 82 mass-richness parameters for all situations.]{Comparison of Stripe 82 mass-richness parameters for all different systematic corrections. The upper four plots compare the mass coefficients and the lower four plots compare the power law slopes. The situations are:  1=No corrections; 2=Photo-z bias correction; 3=BCG correction; 4=Halo miscentering correction; 5=Photo-z uncertainty correction; 6=Foreground galaxy correction; 7=All corrections. Note that the final (all corrections) result from this paper is the large blue circle shown as situation 7.} \label{compareallvalues} 
\end{figure}

	In Figure \ref{compareallvalues} we compare all of the values of the mass coefficient and the power law slope from Table \ref{massrich1}.   
The situations in these plots are:
(1) no corrections; (2) photo-z bias correction; (3) central BCG correction; (4) halo miscentering correction; (5) photo-z uncertainty correction; (6) foreground galaxy correction; and (7) all corrections. The correction that typically has the most significant effect on the mass coefficient and the power law slope is the correction for halo miscentering. Mass coefficient also changes significantly when the five corrections are all applied together. 

	\subsubsection{Comparison to Other Groups}  \label{others}
	
	Directly comparing mass coefficients and power law slopes to other groups is difficult because other samples have a number of elements that make them different. Our measurements are different from those of \citet{Johnston07} and \citet{Simet11} because they use the richness measure $N_{200}$ while we use $N_{VT}$. The Voronoi tessellation cluster sample will have different properties than the maxBCG sample used by these groups as the former is independent of galaxy color while the latter relies on it. Similarly the 3D-Matched-Filter (3D-MF) cluster-finding algorithm used by \citet{Ford14a} and \citet{Ford14b} is independent of galaxy color, but its method of searching for galaxy clusters based on a fiducial luminosity profile will produce a sample with different properties than the VT cluster finder. For example, the 3D-MF cluster finder produces a cluster sample with a much larger richness range than the VT cluster finder. \citet{Ford14b} discuss the fact that their mass coefficient is quite low compared to other measurements; they consider this to be because their cluster finder locates more low-mass clusters than other cluster finders or because it detects more faint galaxies, leading to higher richness values. \citet{Reyes09} use the same sample as \citet{Johnston07} but they define $M_{200}$ based on $\overline{\rho}$ (mean density of the Universe) rather than $\rho_{crit}$ (critical density of the Universe). (\citet{Ford14b} and \citet{Simet11} use critical density in their definitions of $M_{200}$.). Since mean density is slightly larger than critical density, we expect that the mass values found by Reyes would be slightly larger; \citet{Reyes09} predict that mass values found using mean density will be $30\%$ higher. Since \citet{Johnston07} give a mass coefficient of $0.88\times10^{14} \; h^{-1} M_{\sun}$ and \citet{Reyes09} give one of {$1.42\times10^{14} M_{\sun}$ ($61\%$ difference)}, these values approximately follow the expected trend.   
	
	In order to make an approximate comparison, we use Equation \ref{convertNVT} to find that when {$N_{200}=20$, $N_{VT}=12.5$}. Thus we find mass coefficients at $N_{VT}=12.5$, approximating that these correspond to $N_{200}=20$. For the low-z bin we find {$M_{200|N_{200}=20}=5.35 \pm 0.299 \times10^{13} \; h^{-1} M_{\sun}$}. This is smaller than the previous measurements, including $M_{200|20}=8.8 \pm 1.2 \times10^{13} \; h^{-1} M_{\sun}$ from Johnston et al., $M_{200|20}=9.56 \pm 0.75 \times10^{13} \; h^{-1} M_{\sun}$ from Simet et al and $M_{200|20}=14.2 \pm 0.88 \times10^{13} M_{\sun}$ from Reyes et al. For the mid-z bin we found $M_{200|N_{200}=20}=8.68 \pm 0.835 \times10^{13} \; h^{-1} M_{\sun}$. In this approximate redshift range \citet{Ford14a,Ford14b} and found the significantly smaller values of $M_{200|20}=2.2 \pm 0.2 \times10^{13} M_{\sun}$ or $M_{200|20}=2.7^{+0.5}_{-0.4}  \times10^{13} M_{\sun}$. However, as mentioned above, \citet{Ford14a,Ford14b} note that their values of mass coefficient are lower than others.
	
	If we directly compare our values of power law slope to those of other groups, we find that our value for the redshift $0.1-0.4$ bin of {$0.905 \pm 0.0585$} is close to the values found by \citet{Simet11} and \citet{Reyes09} of $1.10 \pm 0.12$ and $1.16 \pm 0.09$. It is farther from the value of \citet{Johnston07} of $1.28 \pm 0.04$. Our values of power law slope for the redshift $0.1-1.0$ bin of {$0.883 \pm 0.0500$} are lower than those found by \citet{Ford14a} and \citet{Ford14b}, which were $1.5 \pm 0.1$ and $1.4 \pm 0.1$. 

\section{Redshift Evolution}  \label{seven}
\begin{figure}  
\begin{subfigure}
\centering
\includegraphics[scale=0.4, angle=90]{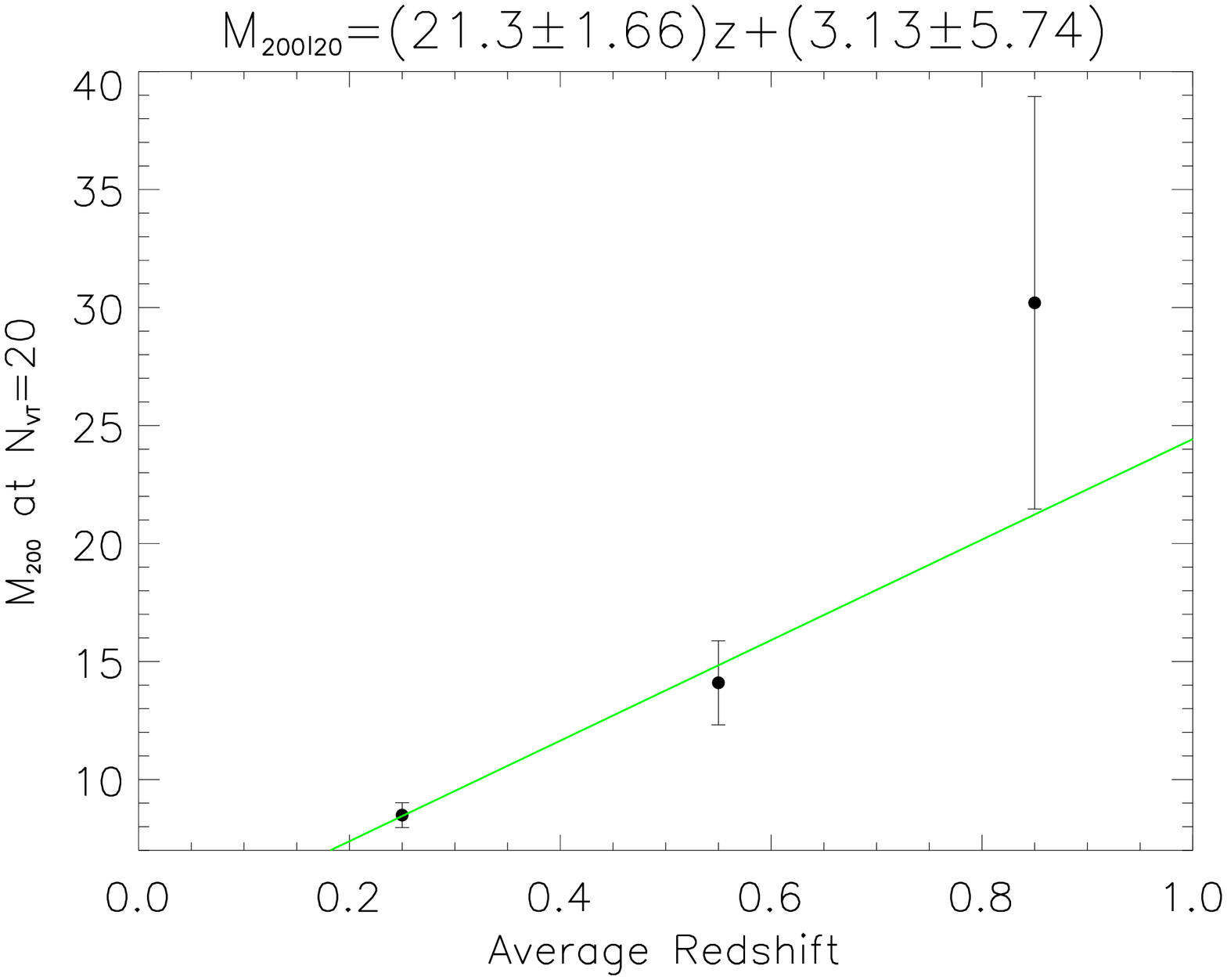}
\end{subfigure}  \\
\begin{subfigure} 
\centering
\includegraphics[scale=0.4, angle=90]{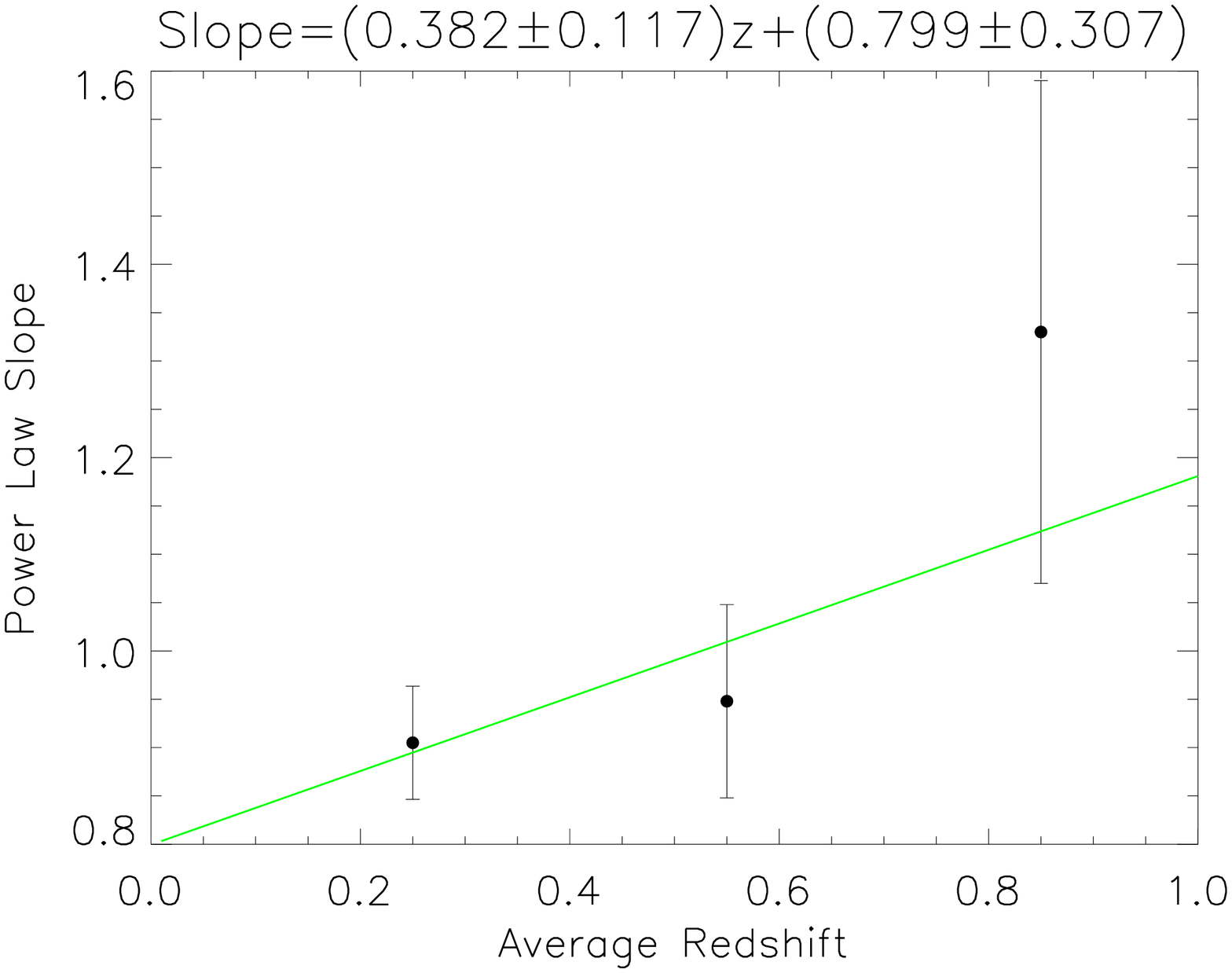}
\end{subfigure} 
\caption[The mass-redshift and power law slope-redshift relation for the Stripe 82 co-add.]{The mass at $N_{VT}=20$ as a function of redshift (upper plot) and the power law slope as a function of redshift (lower plot) for the Stripe 82 co-add. The green lines are the linear best-fitting relations, with their respective equations displayed at the top of each plot. \label{massred}}
\end{figure}

	To examine the evolution of the mass-richness relation, we considered the mass coefficient and the power law slope as a function of redshift. In Figure \ref{massred} we plot the mass coefficient and the power law slope vs. redshift for three redshift bins:  low-z ($\overline{z}=0.25$), mid-z ($\overline{z}=0.55$) and high-z ($\overline{z}=0.85$). 
We then found linear relations between mass coefficient ($M_{200|N_{VT}=20}$) and z and power law slope, ($\alpha$) and z as:
\begin{equation}
\begin{array}{l}  \label{yesme}
M_{200|N_{VT}=20}(z)=(21.3 \pm 1.66)z+(3.13 \pm 5.74) \\ \\
\alpha(z)=(0.382 \pm 0.117)z+(0.799 \pm 0.307)
\end{array}
\end{equation}
Figure \ref{massred} suggests that there is strong evidence of redshift evolution in the mass coefficient. The high-z point has large error bars, but the low-z and mid-z points have smaller error bars and also give evidence of redshift evolution, leading the overall error on the slope of the mass coefficient-redshift relation (Equation \ref{yesme}) to be less than $10\%$ of the slope. Our finding here disagrees with the finding of \citet{Ford14b}:  Figure 8 in \citet{Ford14b} shows the opposite trend, of an overall decrease in mass coefficient with redshift. However their error bars are large and they state that their negative slope is consistent with zero. We find here that redshift evolution of the power law slope is less clear. { While there is some evidence} of redshift evolution of the power law slope, the error on the slope of the second relation in Equation \ref{yesme} is about $30\%$. 

	In Figure \ref{massred} it does appear that there is redshift evolution of the mass coefficient and there may be redshift evolution of the power law slope. This would suggest that clusters of the same richness are more massive at higher redshift {and that at higher redshift, cluster mass} increases faster with increasing richness. This would be consistent with the idea that dark matter haloes associated with galaxy clusters contain more dark matter at higher redshift. However it is important to note that there is a difference between real mass evolution with redshift and apparent mass evolution caused by measurement systematics. One particular systematic that might cause the appearance of redshift evolution is the limiting magnitude for cluster finding. The Voronoi tessellation method uses an apparent magnitude cut rather than an absolute magnitude cut. This means that the sample of clusters will be less complete at higher redshift, where galaxies will be biased toward dimmer apparent magnitude simply because of their large distances. In our sample there are more clusters found at higher redshift than at lower redshift because there is a larger volume sampled at higher redshift. However at increasing redshift, we expect the detection completeness of the cluster sample to decrease. Thus there would be poorer statistics at higher redshift, which could bias the ultimate measurement of mass coefficient at higher-z. In order to further test this mass-redshift relation, it would be better to also find this relation using a cluster sample that is limited by absolute magnitude rather than apparent magnitude. 
	\\
	\section{Conclusions} \label{eight}
		We present mass-richness relations for the Sloan Digital Sky Survey Stripe 82 co-add. These mass-richness relations are presented for four redshift bins, $0.1 < z \leq 0.4$, $0.4 < z \leq 0.7$, $0.7 < z \leq 1.0$ and $0.1 < z \leq 1.0$. We present a sample of galaxy clusters and we describe how these clusters were found using a Voronoi tessellation cluster finder; cluster richness is given in terms of a richness measure we call $N_{VT}$. Our sample includes $19,316$ clusters in Stripe 82 between $0.1 \leq z \leq 0.98$ and {$1 \leq N_{VT} \leq 99$}. We describe how we measured stacked weak lensing shear in a sequence of redshift and richness bins. We found values for $M_{200}$ by fitting measured shear to an NFW model. 
	
	We describe tests conducted to verify the strength of the weak lensing signal; these included measurements of orthotangential shear for all cluster bins and measurements of weak lensing shear around random points. We also describe the effects of systematics on weak lensing shear and thus cluster mass ($M_{200}$) results. These systematics include a bias in photometric redshifts, the effect of the central BCG, halo miscentering, photometric redshift uncertainty and foreground galaxy contamination. We present methods to correct for each of these systematics while measuring $M_{200}$. We find that the halo miscentering correction has the largest effect on mass measurements while the other corrections have smaller effects. 
	
	We present values for the mass coefficient and the power law slope for mass-richness relations found using power law fits to Stripe 82 data for each of the four redshift bins. After considering all corrections we find values of the mass coefficient ($M_{200|20}$) of {$8.49 \pm 0.526$, $14.1 \pm 1.78$}, $30.2 \pm 8.74$ and $9.23 \pm 0.525 \times 10^{13} \; h^{-1} M_{\sun}$ for each of the four redshift bins respectively. We find values of the power law slope ($\alpha$) of $0.905 \pm 0.0585$, $0.948 \pm 0.100$, $1.33 \pm 0.260$ and $0.883 \pm 0.0500$ respectively. We also considered mass evolution of the mass-richness relation. We conclude that there is good evidence of redshift evolution of the mass coefficient and less clear evidence of redshift evolution of the power law slope. {However this apparent redshift} evolution may be a cluster selection effect.
	
	\section{Acknowledgments}
	Funding for the SDSS and SDSS-II has been provided by the Alfred P. Sloan Foundation, the Participating Institutions, the National Science Foundation, the U.S. Department of Energy, the National Aeronautics and Space Administration, the Japanese Monbukagakusho, the Max Planck Society, and the Higher Education Funding Council for England. The SDSS Web Site is http://www.sdss.org/.

	The SDSS is managed by the Astrophysical Research Consortium for the Participating Institutions. The Participating Institutions are the American Museum of Natural History, Astrophysical Institute Potsdam, University of Basel, University of Cambridge, Case Western Reserve University, University of Chicago, Drexel University, Fermilab, the Institute for Advanced Study, the Japan Participation Group, Johns Hopkins University, the Joint Institute for Nuclear Astrophysics, the Kavli Institute for Particle Astrophysics and Cosmology, the Korean Scientist Group, the Chinese Academy of Sciences (LAMOST), Los Alamos National Laboratory, the Max-Planck-Institute for Astronomy (MPIA), the Max-Planck-Institute for Astrophysics (MPA), New Mexico State University, Ohio State University, University of Pittsburgh, University of Portsmouth, Princeton University, the United States Naval Observatory, and the University of Washington.	

Fermilab is operated by Fermi Research Alliance, LLC under Contract No. DE-AC02-
07CH11359 with the United States Department of Energy.

We are grateful for the comments and suggestions of the anonymous referee.

\label{lastpage}
\end{document}